\documentclass[twocolumn,floatfix,pra]{revtex4-1}
\usepackage[utf8]{inputenc}
\usepackage{amsmath}
\usepackage{amssymb}
\usepackage{physics}                                
\usepackage{graphicx}
\usepackage{microtype}                              
\usepackage{newtxtext}                              
\usepackage{newtxmath}                              
\usepackage{dsfont}                                 
\usepackage{color}
\usepackage{hyperref}
\usepackage[export]{adjustbox}
\usepackage[caption=false]{subfig}
\usepackage[scr=boondox,scrscaled=1.05]{mathalfa}   
\RequirePackage{fix-cm}

\DeclareMathSizes{10}{10}{7}{5}

\newcommand{\mc}[1]{\mathcal{#1}}                   
\DeclareMathOperator{\Span}{span}

\hypersetup{
    colorlinks = true,
    linkcolor  = blue,
    citecolor  = blue,
    urlcolor   = blue,
}

\begin{document}

\title{Entanglement negativity and sudden death in the toric code at finite temperature}
\author{O.~Hart and C.~Castelnovo}
\affiliation{T.C.M.~Group, Cavendish Laboratory,  J.J.~Thomson Avenue, Cambridge CB3 0HE, United Kingdom}
\date{April 2018}

\begin{abstract}
    We study the fate of quantum correlations at finite temperature in the two-dimensional toric code using the logarithmic entanglement negativity. We are able to obtain exact results that give us insight into how thermal excitations affect quantum entanglement.
    The toric code has two types of elementary excitations (defects) costing different energies.
    We show that an $\mc{O}(1)$ density of the lower energy defect is required to degrade the zero-temperature entanglement between two subsystems in contact with one another. However, one type of excitation alone is not sufficient to kill all quantum correlations, and an $\mc{O}(1)$ density of the higher energy defect is required to cause the so-called sudden death of the negativity. Interestingly, if the energy cost of one of the excitations is taken to infinity, quantum correlations survive up to arbitrarily high temperatures, a feature that is likely shared with other quantum spin liquids and frustrated systems in general, when projected down to their low-energy states. 
    We demonstrate this behaviour both for small subsystems, where we can prove that the negativity is a necessary and sufficient condition for separability, as well as for extended subsystems, where it is only a necessary condition. We further observe that the negativity per boundary degree of freedom at a given temperature increases (parametrically) with the size of the boundary, and that quantum correlations between subsystems with extended boundaries are more robust to thermal fluctuations.
\end{abstract}

\maketitle


\section{introduction}

With the modern surge of interest in harvesting the capabilities of quantum mechanical systems to develop new technologies, it has become ever so important to quantify and characterise quantum correlations in physical systems. While this remains a tall order in real systems, substantial progress has been made in recent years at the theoretical level. Several measures of quantum correlations (including, but not limited to, ``entanglement’') have been proposed and studied (see Refs.~\onlinecite{Plenio2007, Amico2008, Laflorencie2016} for reviews). One of the most successful is the von Neumann entanglement entropy. Similarly to other measures, the entanglement entropy works remarkably well when a system is prepared in a pure state---namely, when the density matrix is a projector onto a single quantum mechanical state. However, it becomes less descriptive once we deal with mixed states, as is the case for systems at finite temperature. Under these circumstances, the von Neumann entropy becomes dominated by classical correlations~\cite{Sherman2016} and it is difficult to distil quantum from classical correlations.

One of the most promising ways to address this issue has come of late from a newly introduced measure of entanglement, the so called (logarithmic) \emph{entanglement negativity} $\mc{E}$~\cite{Karol1998, Vidal2002} and variants thereof~\cite{Shapourian2017, Eisert2016}. The entanglement negativity uses the positive partial transpose criterion~\cite{Peres1996} to detect the separability of a system's density matrix, and as such it is a proper measure of purely quantum correlations---although it is in general only a necessary but not sufficient condition for separability~\cite{Eisert2001, Shapourian2017}.

Unfortunately, calculating $\mc{E}$ in many-body quantum systems is notoriously difficult and analytic results are few. At zero temperature, the negativity has been evaluated for a limited number of sufficiently simple lattice models in one~\cite{Wichterich2010, Santos2011, Santos2016} and two~\cite{Castelnovo2013, Lee2013} dimensions, suggesting universality at quantum critical points and unveiling leading area law behaviour respectively. That the negativity is fully universal and scale invariant at quantum critical points has subsequently been proven rigorously using conformal field theory (CFT)~\cite{Calabrese2012, CalabreseCardy2013}. When temperature is raised from zero, one expects that thermal mixing of eigenstates (``thermal fluctuations'') will lead to a reduction in the entanglement. This intuition has been verified for a large number of \emph{two}-spin systems~\footnote{In fact, these studies show that entanglement is \emph{not} always a monotonic function of temperature. Particularly in the presence of a large magnetic field, it is possible for entanglement, as measured by the concurrence, to initially \emph{increase} with increasing temperature before eventually vanishing at $T_c$, see e.g.~Refs.~\onlinecite{Arnesen2001, Mehran2014}.}, initially using the concurrence~\cite{Gunlycke2001, Osborne2002, Arnesen2001, Wang2001, Wang2002, Wang2002Sep, Zhang2005, Li2008, Zhou2003, Rigolin2004, Lagmago2002, Mehran2014} (for a review, see Ref.~\onlinecite{Amico2008}) and more recently using the negativity~\cite{Wang2006, Xiao-San2009, Xu2013}. Generically, there exists a well-defined temperature $T_c$, dubbed the ``sudden death temperature,'' above which the negativity vanishes identically. Beyond two-spin systems, in particular considering the entanglement between two blocks of spins, far less is known. At quantum critical points, CFT results have been extended into the regime of finite temperature~\cite{Eisler2014, Calabrese2015}, while on the numerical side a linked cluster expansion has been employed, alongside exact diagonalisation, to study the negativity at finite temperature in one- and two-dimensional bipartite spin systems~\cite{Sherman2016}. Even though the negativity does not capture \emph{all} quantum correlations, sudden death is nevertheless an intriguing phenomenon that requires further investigation.
Further results in $d > 1$ systems, and in particular exact expressions for the negativity at finite temperature, would be highly beneficial to gain a better understanding about the fate of quantum correlations and the origin of the sudden death behaviour.

The goal of this manuscript is to present the first exact calculation of the negativity in a 2D lattice system in thermal equilibrium at finite temperature.
We choose to work with the toric code model~\cite{Kitaev2003} whose exact solubility has allowed the negativity at $T=0$~\cite{Castelnovo2013, Lee2013} and the von Neumann entropy at finite temperature~\cite{Castelnovo2007} to be calculated exactly.
The dynamical approach to thermal equilibrium, and the resulting relaxation of the negativity, is beyond the scope of this work. A review of quantum memories at finite temperature, which highlights the relevant time scales associated with equilibration, can be found in Ref.~\onlinecite{Brown2016}.
We consider the entanglement between two subsystems that share a boundary of finite length, which allows us to discern the fate of the zero-temperature area law, and how the sudden death temperature $T_c$ depends on boundary length.

We find that the zero-temperature area law holds for all temperatures, unlike the von Neumann entropy which transitions to a volume law for sufficiently large temperatures~\cite{Castelnovo2007, Eisert2010}. We also show that, \emph{irrespective of boundary length}, the negativity is killed by an $\mc{O}(1)$ density of the most energetically costly defect. The entanglement between larger partitions is however more robust to the effects of thermal fluctuations, with the sudden death temperature of the largest boundaries being approximately twice that of the smallest boundaries. Sending the energy cost of the most expensive defects to infinity removes sudden death, and instead a gradual $1/T$ demise of the negativity is observed.

The manuscript is arranged as follows. We briefly introduce the toric code model, the entanglement negativity, and the techniques pertaining to its calculation in Sec.~\ref{sec:introduction_to_negativity} and Sec.~\ref{sec:introduction_to_model}. A summary of our main (exact) analytical results, as well as a discussion of their significance, is presented in Sec.~\ref{sec:main_results}. Details of the relatively long and technical analytical calculations are for convenience postponed until Sec.~\ref{sec:calculations}.
We then briefly comment on the potential experimental relevance of our results in Sec.~\ref{sec:experimental_considerations}.
Finally, in Sec.~\ref{sec:conclusions}, we draw our conclusions and outlook.


\subsection{Entanglement negativity}
\label{sec:introduction_to_negativity}

Given a tripartite system $\mc{S} = \mc{A}_1 \cup \mc{A}_2 \cup \mc{B}$, the logarithmic negativity is defined in terms of the reduced state $\rho_\mc{A} = \Tr_\mc{B} \rho$ as
\begin{equation}
    \mc{E} = \ln \norm\big{\rho_\mc{A}^{T_2}}_1
		\, ,
\end{equation}
where $T_2$ denotes partial transposition over the $\mc{A}_2$ subsystem and $\norm{\, \cdot \,}_1 = \Tr\abs{\, \cdot \,}$ is the trace norm~\cite{Vidal2002}. The logarithmic negativity quantifies the entanglement between the subsystems $\mc{A}_1$ and $\mc{A}_2$. One can verify that it is symmetric in $\mc{A}_1 \leftrightarrow \mc{A}_2$ as a good measure of entanglement must be, so that we can equally compute $\norm\big{\rho_\mc{A}^{T_1}}_1$.

Given an (arbitrary) orthonormal basis $\ket*{\psi_i \phi_j} \equiv \ket*{\psi_i} \otimes \ket*{\phi_j} \in \mc{H}_{\mc{A}_1} \otimes \mc{H}_{\mc{A}_2}$, the operation of partial transposition can be defined in terms of matrix elements as~\cite{Vidal2002}
\begin{equation}
    \mel*{\psi_i \phi_j}{\rho_\mc{A}^{T_2}}{\psi_k \phi_l} = \mel*{\psi_i \phi_l}{\rho_\mc{A}}{\psi_k \phi_j}
		\, .
\end{equation}

In order to evaluate the negativity analytically, we employ the replica approach, introduced by the authors of Ref.~\onlinecite{Calabrese2012}, which has since been used to calculate the negativity in a variety of models~\cite{Calabrese2012, Castelnovo2013, CalabreseCardy2013, Calabrese2013, Calabrese2015, Ruggiero2016, Ruggiero2016}. If the operator $ \rho_\mc{A}^{T_2} $ has eigenvalues $\lambda_i$, then
\begin{align}
    \Tr\big(\rho_\mc{A}^{T_2}\big)^{n_\text{e}} &= \sum_{\lambda_i \geq 0} \abs*{\lambda_i}^{n_\text{e}} + \sum_{\lambda_i < 0} \abs*{\lambda_i}^{n_\text{e}} 
		\, , \\
    \Tr\big(\rho_\mc{A}^{T_2}\big)^{n_\text{o}} &= \sum_{\lambda_i \geq 0} \abs*{\lambda_i}^{n_\text{o}} - \sum_{\lambda_i < 0} \abs*{\lambda_i}^{n_\text{o}}
		\, ,
\end{align}
where $n_\text{e}$ ($n_\text{o}$) is an even (odd) positive integer. Crucially, in contrast to $\rho_\mc{A}$, the operator $\rho_\mc{A}^{T_2}$ can have negative eigenvalues. To obtain the sum of the absolute values of the eigenvalues, i.e.,~the trace norm, the replica method consists of following the \emph{even} series and taking the analytic continuation $n_\text{e} \to 1$. In this way, we are able to evaluate the negativity exactly via the expression
\begin{equation}
    \mc{E} = \lim_{n_\text{e} \to 1} \ln \Tr\big(\rho_\mc{A}^{T_2}\big)^{n_\text{e}}
		\, . 
		\label{eqn:replica_approach_expression}
\end{equation}
%
%


\subsection{Toric code model}
\label{sec:introduction_to_model}

The toric code is defined on a 2D square lattice composed of $N$ sites with periodic boundary conditions (i.e., on a torus). Spin-$\tfrac12$ degrees of freedom are located on each of the $2N$ bonds of the lattice and interact via the Hamiltonian
\begin{equation}
    H = -\lambda_A \sum_s A_s - \lambda_B \sum_p B_p 
		  \equiv - \lambda_A S - \lambda_B P
		\, ,
    \label{eqn:toric_code_hamiltonian}
\end{equation}
where the labels $s$ and $p$ denote the ``stars'' and ``plaquettes'' of the lattice, respectively, and the operators $A_s \equiv \prod_{i\in s} \sigma_i^x$, $B_p \equiv \prod_{i\in p} \sigma_i^z$~\cite{Kitaev2003}. $\lambda_A, \lambda_B > 0$ are the two coupling constants of the model. As usual $\boldsymbol{\sigma}_i=(\sigma_i^x, \sigma_i^y, \sigma_i^z)$ denote the Pauli matrices, which describe the spin on bond $i$.

\begin{figure}[b]
    \centering
    \includegraphics[width=0.65\linewidth]{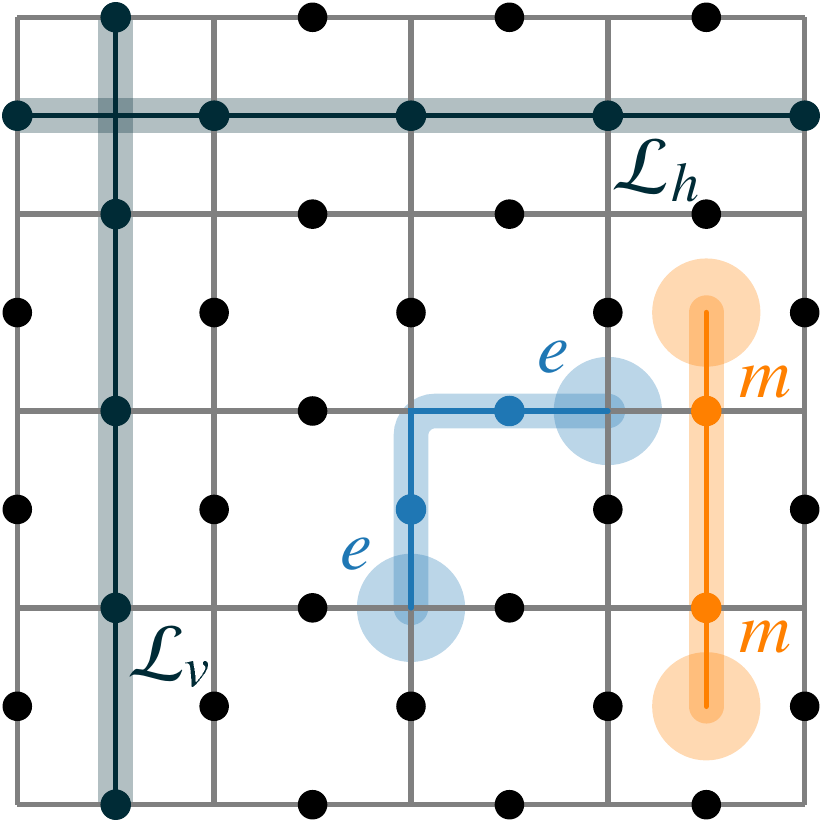}
    \caption{Illustration of two possible noncontractible loops $\mc{L}_{h,v}$, and of two open strings generating electric ($e$) and magnetic ($m$) defects at their ends. Recall that periodic boundary conditions are imposed. Each electric (magnetic
    ) defect costs an energy $2\lambda_A$ ($2\lambda_B$) so that each open string costs $4\lambda_A$ ($4\lambda_B$).}
    \label{fig:toric_code_lattice}
\end{figure}

All operators $A_s$, $B_p$ in the Hamiltonian mutually commute and so can be diagonalised simultaneously. The property $A_s^2 = \mathds{1} = B_p^2$ $\forall$ $s, p$ implies that the operators $A_s$ and $B_p$ have eigenvalues $\pm 1$, and the states for which all $A_s$, $B_p$ have eigenvalue $+1$ are the ground states. The model has 4 ground states belonging to different ``topological sectors,'' which are classified according to the eigenvalues of nonlocal (system-spanning) operators. Such operators can be defined for instance as $\Gamma_{h,v}^{x} = \prod_{i \in \mc{L}_{h,v}} \sigma_i^{x}$~\cite{Trebst2007}, where $\mc{L}_{h,v}$ are two noncontractible loops on the dual lattice that span the torus in the horizontal ($h$) and vertical ($v$) directions, respectively (see Fig.~\ref{fig:toric_code_lattice}).

There exist two types of elementary excitations: \emph{electric charges} and \emph{magnetic vortices} generated by open strings of Pauli matrices, $S^\alpha(\ell)=\prod_{i\in\ell}\sigma_i^\alpha$ with $\alpha = z, x$ respectively~\cite{Carr2010} (see Fig.~\ref{fig:toric_code_lattice}). $S^\alpha(\ell)$ flips the eigenvalues of the \emph{two} stars (plaquettes) at the ends of the path $\ell$, which lives on the direct (dual) lattice, costing an energy $4\lambda_A$ ($4\lambda_B$). With periodic boundary conditions, it is not possible to change the eigenvalue of just one star or plaquette operator as a result of the condition $\prod_s A_s = \prod_p B_p = \mathds{1}$. At finite temperature, excitations are present in the system with a density given approximately by the Fermi-Dirac distribution $n_\text{F}(2\lambda_X) \equiv (\mathrm{e}^{2\beta\lambda_X} + 1)^{-1}$, for $X = A, B$, in the limit of large system size $N$ (see Appendix~\ref{sec:defect_density} for further details).


\section{main results}
\label{sec:main_results}

\begin{figure}[b]
    \subfloat[]{%
        \includegraphics[width=0.3\linewidth, valign=c]{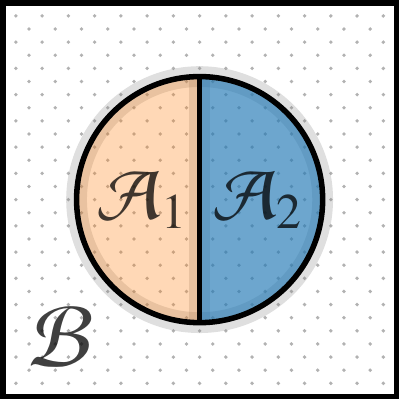}
        \label{fig:partitions:tripartite}
    }%
    \hspace{0.15\linewidth}%
    \subfloat[]{%
        \includegraphics[width=0.3\linewidth, valign=c]{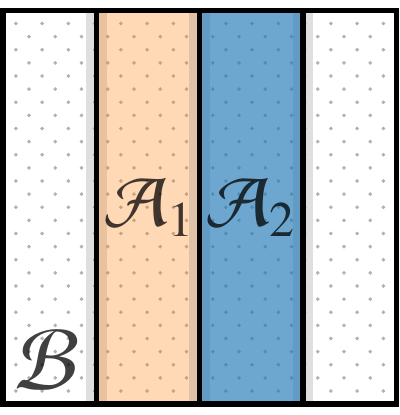}
        \label{fig:partitions:strip}
    }%
    \caption{Schematic depiction of the two types of partitions considered in this manuscript. Note that in (a) the subsystem $\mc{A} = \mc{A}_1 \cup \mc{A}_2$ is contractible, while in (b) it is not. As a result, we expect to see a constant topological contribution to the zero-temperature negativity when using partition scheme (b)~\cite{Castelnovo2013, Lee2013}.}
    \label{fig:partitions}
\end{figure}

We consider a tripartition of the system $\mc{S}$ into three disjoint subsystems $\mc{S} = \mc{A}_1 \cup \mc{A}_2 \cup \mc{B}$. Our goal is to quantify the entanglement between subsystems $\mc{A}_1$ and $\mc{A}_2$ using the logarithmic negativity $\mc{E}$, when the system is in a thermal state $\rho = \mathrm{e}^{-\beta H} / Z$. To obtain explicit expressions, it is necessary to make some assumptions about the arrangement of the various subsystems. The partition schemes with which we choose to work are shown schematically in Fig.~\ref{fig:partitions}. In all cases, subsystems $\mc{A}_1$ and $\mc{A}_2$ have one common edge since it is known for the toric code that the zero-temperature negativity vanishes identically if the subsystems are spatially separated~\cite{Castelnovo2013, Lee2013}. The length of this boundary is parameterised by $N_\partial$, which we define as the number of stars that straddle the $\mc{A}_1$--$\mc{A}_2$ boundary (hereinafter ``boundary stars''). In Fig.~\ref{fig:partitions:strip}, where subsystem $\mc{A} =\mc{A}_1 \cup \mc{A}_2$ spans the system in one direction, we expect to obtain a constant, additive contribution to the zero-temperature negativity, indicative of topological order~\cite{Castelnovo2013}.

\subsection{Star plaquette pair}
\label{sec:single_star_system}

It is useful to study first the smallest subsystem $\mc{A}$ that gives a nonvanishing negativity: a neighbouring star plaquette pair (SPP), embedded in an arbitrarily large lattice of $2N$ spins. This system choice allows us to gain some intuition about the behaviour of the negativity at finite temperatures without any of the additional complications presented by more general partition schemes. The fact that we cannot simply consider a neighbouring pair of spins is a direct consequence of the four-body interactions present in the Hamiltonian~\eqref{eqn:toric_code_hamiltonian}. The SPP system offers a number of technical advantages. First, the negativity $\mc{E}$ is both a necessary and sufficient condition for separability, which we prove in Sec.~\ref{sec:single_star_calculation}. This allows us to study the entanglement properties of the system exactly. Also, the SPP is symmetric under interchange of stars with plaquettes, implying that the negativity must obey the symmetry $\mc{E}(\lambda_A, \lambda_B) = \mc{E}(\lambda_B, \lambda_A)$. Finally, one can diagonalise the operator $\rho_\mc{A}^{T_2}$ explicitly without having to resort to the replica trick. A short calculation, outlined in Sec.~\ref{sec:single_star_calculation}, gives
\begin{equation}
    {\mc{E}(T)} = \ln\left[1 + \tfrac12 \max(\eta_A + \eta_B + \eta_A\eta_B - 1, 0)\right]
        \, ,
    \label{eqn:single_star_negativity}
\end{equation}
where we have defined
\begin{equation}
    \eta_X = \frac{\tanh(\beta\lambda_X) + \tanh(\beta\lambda_X)^{N-1}}{1 + \tanh(\beta\lambda_X)^N}
        \, ,
    \label{eqn:eta_redefined}
\end{equation}
for $X=A,B$. The expression~\eqref{eqn:single_star_negativity} manifestly respects the symmetry under interchange of $\lambda_A$ with $\lambda_B$. Since $\eta_X$ is a monotonic increasing function of $\beta$, the negativity is a monotonically decreasing function of temperature, with a single discontinuity in its first derivative at the sudden death temperature $T_c$. Above $T_c$, the state of subsystem $\mc{A}$ is separable. Agreement with numerical exact diagonalisation of $\rho_\mc{A}^{T_2}$ is perfect (not shown).

Recall that when $N \gg 1$, the defect density of species $X$, when the system is in thermodynamic equilibrium, is approximately given by the Fermi-Dirac distribution $n_\text{F}(2\lambda_X) = (\mathrm{e}^{2\beta\lambda_X} + 1)^{-1}$.
The condition for sudden death is then written most succinctly in terms of the density of \emph{holes} $n_h(2\lambda_X) \equiv 1 - n_\text{F}(2\lambda_X)$, i.e.,~the density of nondefective stars (plaquettes) for $X=A \, (B)$.
In the thermodynamic limit, this condition becomes
\begin{equation}
    n_h(2\lambda_A) n_h(2\lambda_B) = \frac{1}{2}
        \, .
    \label{eqn:sudden_death_condition_physical}
\end{equation}
That is, at $T_c$, the geometric mean of the hole densities must equal $1/\sqrt{2}$. When the coupling constants are well separated, one can obtain an approximate explicit expression for the critical temperature
\begin{equation}
    \frac{T_c}{\lambda_A} \simeq
        \dfrac{2\frac{\lambda_B}{\lambda_A}}{W\left(2\frac{\lambda_B}{\lambda_A}\right)} \sim \dfrac{2\frac{\lambda_B}{\lambda_A}}{\ln\left(2\frac{\lambda_B}{\lambda_A}\right)} \quad \text{ for } \quad \lambda_B \gg \lambda_A
            \, ,
    \label{eqn:SPP_sudden_death_limit}
\end{equation}
where $W(x)$ is the product-log function. Exploiting the duality, the expression for $\lambda_A \gg \lambda_B$ is found by exchanging $\lambda_A \leftrightarrow \lambda_B$. The energy scale for sudden death is set by the larger of the two coupling constants, closely mirroring the perturbative expressions of Ref.~\onlinecite{Sherman2016}. In the classical limit $\lambda_A \to 0$, keeping $\lambda_B$ fixed, $T_c$ vanishes logarithmically.

\begin{figure}
    \centering
    \includegraphics[width=0.9\linewidth]{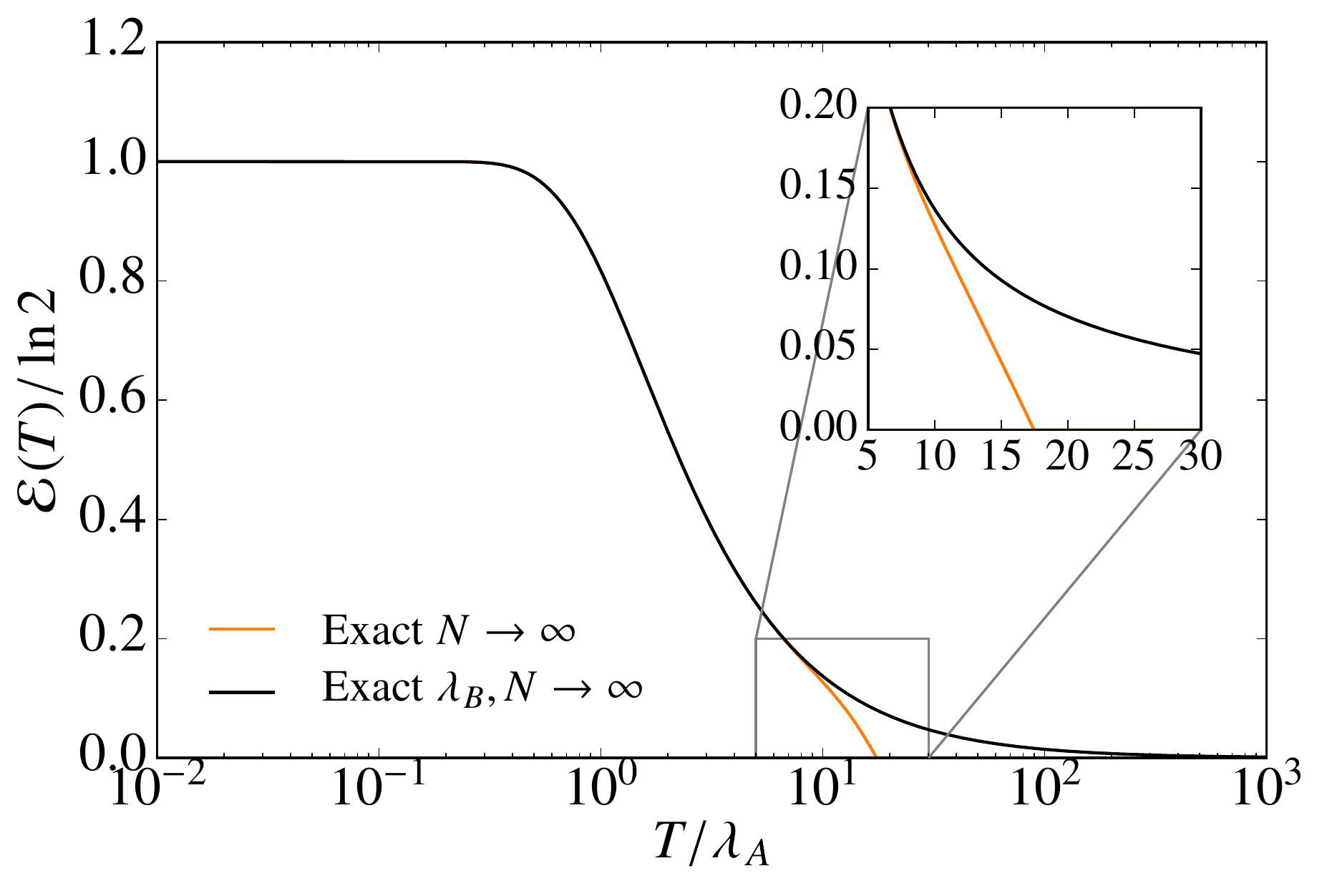}
    \caption{The analytic result~\eqref{eqn:single_star_negativity}, and its limit with $\lambda_B \to \infty$~\eqref{eqn:tripartite_negativity_final_thermodynamic}, for the SPP embedded in an infinite system $N\to \infty$ with $\lambda_B / \lambda_A = 25$. The $\lambda_B \to \infty$ limit provides an excellent approximation up to a temperature of order $\lambda_B$, where it becomes possible to thermally excite magnetic vortices and the assumption of undegraded plaquette magnetisation $M(\alpha) = N$ is no longer valid.}
    \label{fig:numerical_comparison_tripartite}
\end{figure}

The result~\eqref{eqn:single_star_negativity} allows for the following physical interpretation. When the temperature is finite but low compared to $\lambda_A$, the densities of both defect species are exponentially suppressed ($n_h \simeq 1 - \mathrm{e}^{-2\beta\lambda_X}$). As can be seen in Fig.~\ref{fig:numerical_comparison_tripartite}, over the temperature range $T \lesssim \lambda_A$, the negativity remains essentially undegraded from its zero-temperature value, $\mc{E}\simeq \mc{E}(0)$. Namely, a density of order one defect per site of the lattice of the least energetically costly defects is needed to start affecting the quantum correlations between $\mc{A}_1$ and $\mc{A}_2$. Above the temperature threshold $T \sim \lambda_A$, electric charge defects proliferate in the system and correspondingly the negativity decays like $\mc{E} \sim \lambda_A / T$. This decay is cut off at a temperature $T \sim \lambda_B$ (up to logarithmic factors), corresponding to an $\mc{O}(1)$ density of the most energetically costly (magnetic) defects in the system, where the negativity vanishes.

\subsection{Extended boundary}
\label{sec:larger_boundary}

The SPP results are simple and intuitive, but do not tell us how the negativity depends on the length of the $\mc{A}_1$--$\mc{A}_2$ boundary.
To address this, we consider the subsystem $\mc{A}$ depicted in Fig.~\ref{fig:partitions:strip}, i.e.,~a ``strip'' that spans the torus in one direction. One can imagine constructing this partition scheme by tiling neighbouring SPPs. Choosing such a noncontractible subsystem makes the calculation slightly simpler than for a strip of finite length, by eliminating its end points (although, as shown in Sec.~\ref{sec:larger_boundary_calculation}, the two results are indeed in agreement).

In the thermodynamic limit $N\to \infty$, keeping the length of the boundary $N_\partial$ finite (namely, for a system wrapped around a cylinder of infinite length), the replica method gives (see Sec.~\ref{sec:larger_boundary_calculation})
\begin{equation}
    \mc{E}(T) = \ln \big\langle \, \big\vert 
			        Z(\{ \gamma_s \}, \{ k_s \}, T) 
			    \big\vert \, \big\rangle
		\, .
    \label{eqn:negativity_larger_boundary}
\end{equation}
Here $Z$ is proportional to the partition function of a disordered 1D Ising chain with periodic boundary conditions in a complex magnetic field,
\begin{equation}
    Z = \frac{1}{(\cosh\beta\lambda_B)^{N_\partial}} \sum_{\{ \tau_s = \pm 1 \}} \mathrm{e}^{-H(\{ \tau_s \})} \, ,
\end{equation}
with the classical ``Hamiltonian''
\begin{equation}
    H(\{ \tau_s \}) = \sum_{s=1}^{N_\partial} \left[ \gamma_s \tau_s \tau_{s+1} + \frac12 (K_A + \mathrm{i}\pi k_s ) (\tau_s + 1) \right]
        \, .
\end{equation}
The angled brackets in Eq.~\eqref{eqn:negativity_larger_boundary} refer to a disorder average over the variables $\{ \gamma_s \}$ and $\{ k_s \}$, with $\gamma_s \in \{ -\beta\lambda_B, \beta\lambda_B \}$ and $k_s \in \{ 0, 1 \}$ with equal probability.

\begin{figure}[t]
    \centering
    \includegraphics[width=0.9\linewidth]{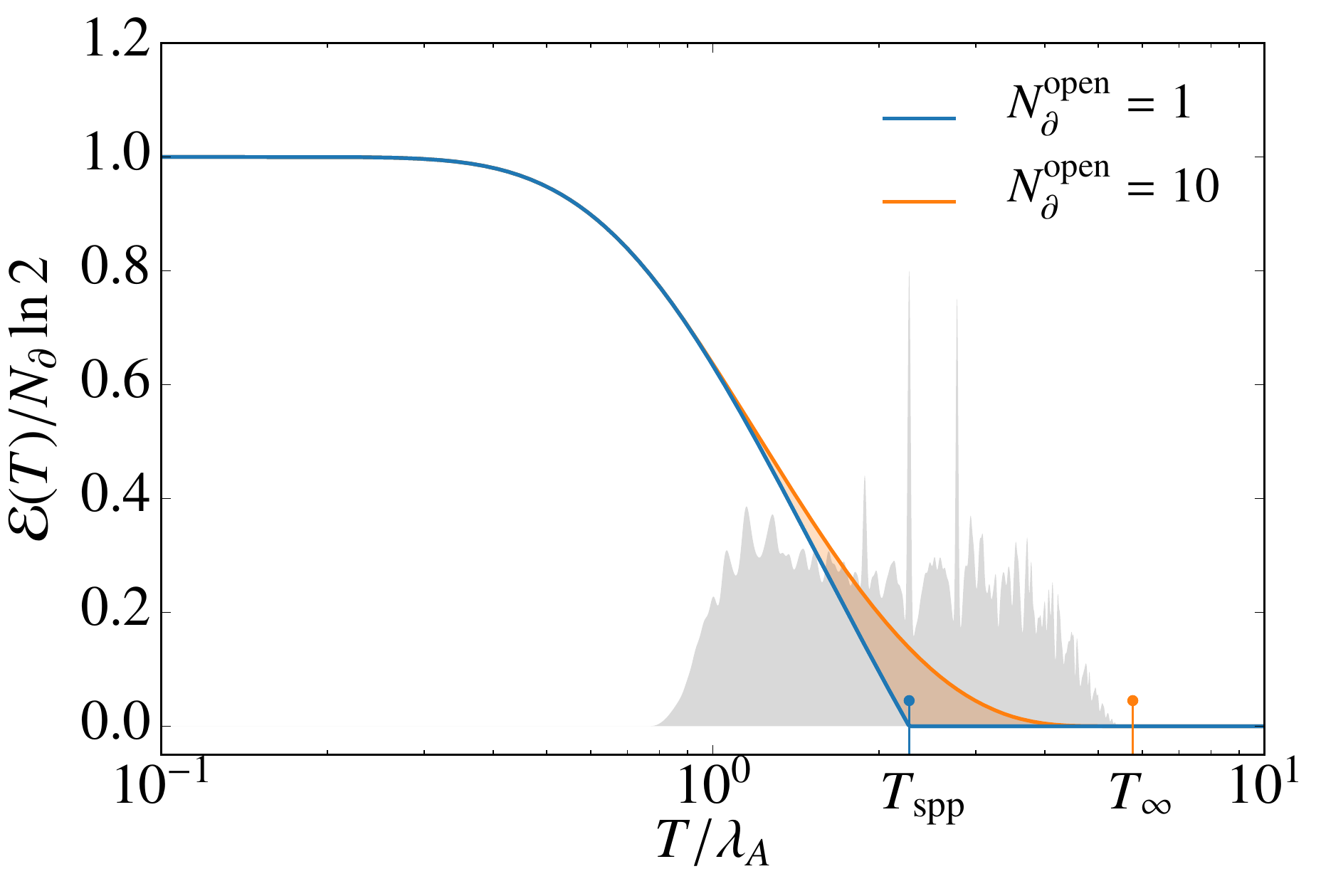}
    \caption{Negativity for a contractible strip partition, as discussed in the text, with boundary length $N_\partial^\text{open}$ and $\lambda_A = \lambda_B$. As the boundary between $\mc{A}_1$ and $\mc{A}_2$ is made longer, $\mc{E}$ develops an increasing number of discontinuities (at temperatures $T_i$) in its first derivative whose effect is to ``smooth'' the function $\mc{E}(T)$ and push $T_c$ to higher temperatures. In the background is plotted the ``density of discontinuities'' $\sum_i \delta(T - T_i)$ for $N_\partial^\text{open}=7$. The vertical lines indicate the positions of $T_c$ for the SPP and for an infinite boundary.}
    \label{fig:boundary_length_equal}
\end{figure}

Note that because the magnetic field is complex, the partition function is not necessarily positive. Details of the derivation are given in Sec.~\ref{sec:larger_boundary_calculation}, and we have confirmed that the finite-size version of Eq.~\eqref{eqn:negativity_larger_boundary} [that is, Eq.~\eqref{eqn:trace_norm_larger_boundary_finitesize} in Sec.~\ref{sec:larger_boundary_calculation}] agrees precisely with numerical exact diagonalisation results for numerically accessible system sizes (not shown).

Because the subsystem $\mc{A}$ spans the torus in one direction, in addition to leading area law behaviour, the zero-temperature negativity $\mc{E}(0) = (N_\partial - 1)\ln 2$ contains a constant topological contribution, consistent with Refs.~\onlinecite{Castelnovo2013, Lee2013}. The reader is referred to Sec.~\ref{sec:larger_boundary_calculation} for a derivation of this result.
At zero temperature, the result remains unchanged for more general partitions in which subsystems $\mc{A}_1$ and $\mc{A}_2$ have larger widths.
This reflects that the boundary is the source of entanglement, and that the toric code is a zero-range correlated model~\cite{Castelnovo2013}.
It is worth noting that although $\lim_{T\to 0} \rho$ is not pure, owing to the ground state degeneracy, each topological sector gives an independent and equal contribution to $\mc{E}(0)$.

One can show that the disorder average without taking the absolute value is identically unity, i.e.,~$\ev*{Z} \equiv 1$, leading to a vanishing negativity. If at least one disorder realisation satisfies $Z(\{ \gamma_s\}, \{ k_s \}) < 0$, then the average $\ev*{\abs*{Z}}>1$ and consequently $\mc{E} > 0$, indicating the presence of entanglement. For any particular disorder realisation, there exists some temperature $T_*(\{ \gamma_s\}, \{ k_s \}) \geq 0$ above which $Z(\{ \gamma_s\}, \{ k_s \}) > 0$ for all temperatures $T \geq T_*$. Such a temperature must exist since $Z \to 1$ for all disorder realisations in the limit of infinite temperature. The largest of these temperatures corresponds to the sudden death temperature $T_c = \max_{\{\gamma_s, k_s \}} T_*$.

As temperature is increased from zero, similarly to the SPP case in Sec.~\ref{sec:single_star_system}, the negativity remains undegraded, $\mc{E} \simeq \mc{E}(0)$, until the density of thermally excited electric defects becomes of $\mc{O}(1)$. It then vanishes identically above some temperature $T_c$, discussed in more depth below. The SPP and extended case however differ in that the latter exhibits a large number of discontinuities. These discontinuities correspond to the (Lee-Yang~\cite{Yang1952, Lee1952}) zeros of the complex partition functions, whose effect is to ``smooth out'' $\mc{E}(T)$, and to push $T_c$ to higher temperatures (see Fig.~\ref{fig:boundary_length_equal}). This number grows rapidly (exponentially for $N_\partial \gg 1$) with increasing boundary length, e.g.,~$1, 19, 78, \ldots$ for $N_\partial = 2, 3, 4, \ldots$ and $\lambda_A = \lambda_B$.

The disorder realisation that maximises $T_*$, i.e.,~the partition function with the largest zero, corresponds to the special ordered configuration of parameters $ \gamma_s = \beta\lambda_B$ and $k_s = 1$, $\forall \, s$. Since the resulting Ising chain is translationally invariant, one can use a transfer matrix approach to calculate $Z$ analytically. The critical temperature $T_c$ is then found by solving $Z=0$, which in general has multiple solutions as the partition function oscillates. The largest one corresponds to $T_c$, while the others correspond to the positions of discontinuities at lower temperatures. Following this argument, one arrives at the following implicit expression for $T_c$:
\begin{equation}
    \mathrm{e}^{4\beta_c(\lambda_A + \lambda_B)} - \mathrm{e}^{4\beta_c\lambda_A} - \mathrm{e}^{4\beta_c\lambda_B} = \tan^2\left(\frac{\pi}{2N_\partial}\right)
        \, ,
    \label{eqn:sudden_death_condition_larger_boundary}
\end{equation}
which manifestly respects the duality in $\lambda_A \leftrightarrow \lambda_B$, as expected. In contrast to the SPP system of Sec.~\ref{sec:single_star_system}, this equation remains the same when $N$ is taken to be finite, and even when subsystems $\mc{A}_1$ and $\mc{A}_2$ have nontrivial widths (not shown).

\begin{figure}[t]
    \centering
    \subfloat[]{%
        \includegraphics[width=0.471\linewidth]{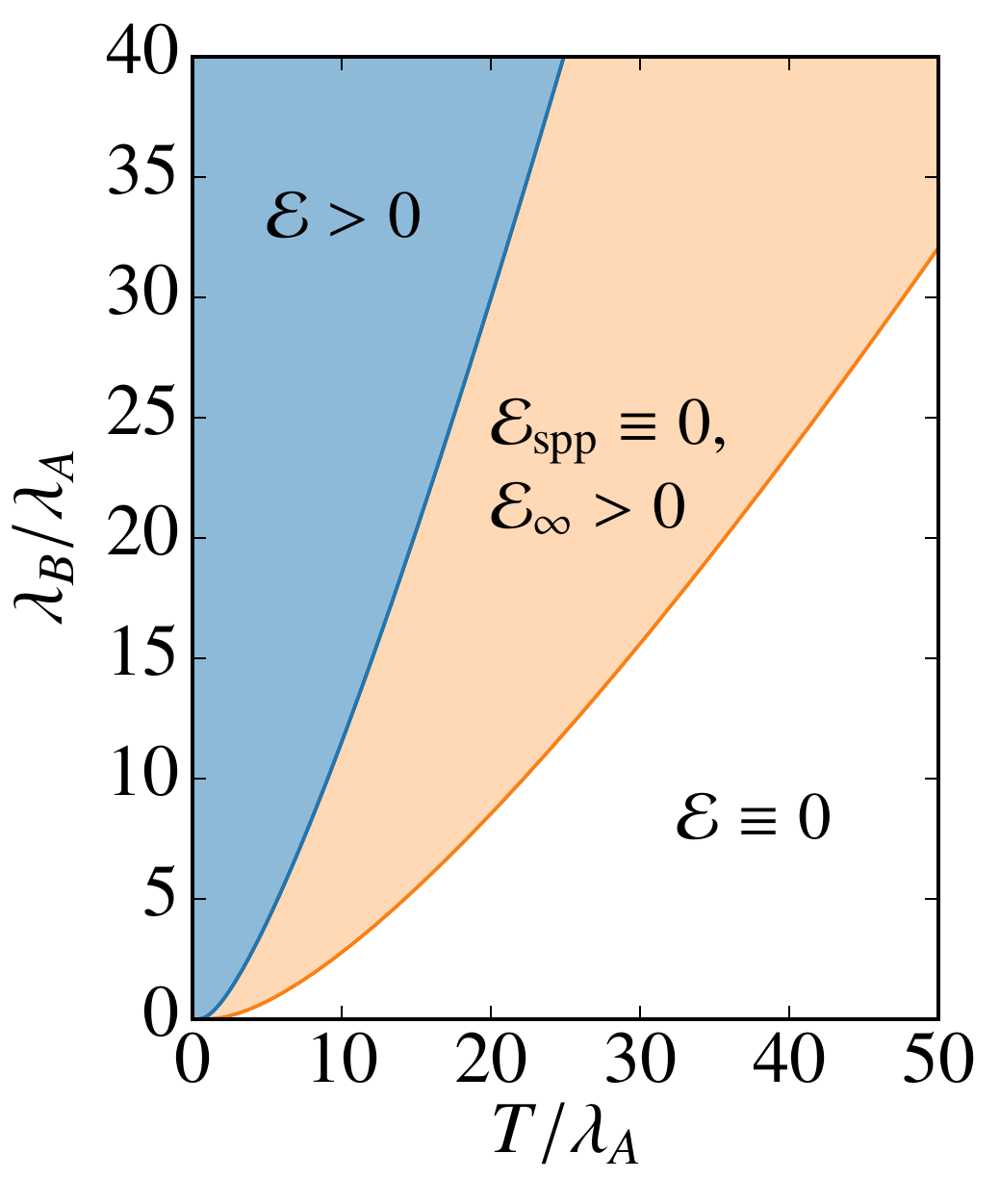}
    }%
    \hfill
    \subfloat[]{%
        \includegraphics[width=0.5\linewidth]{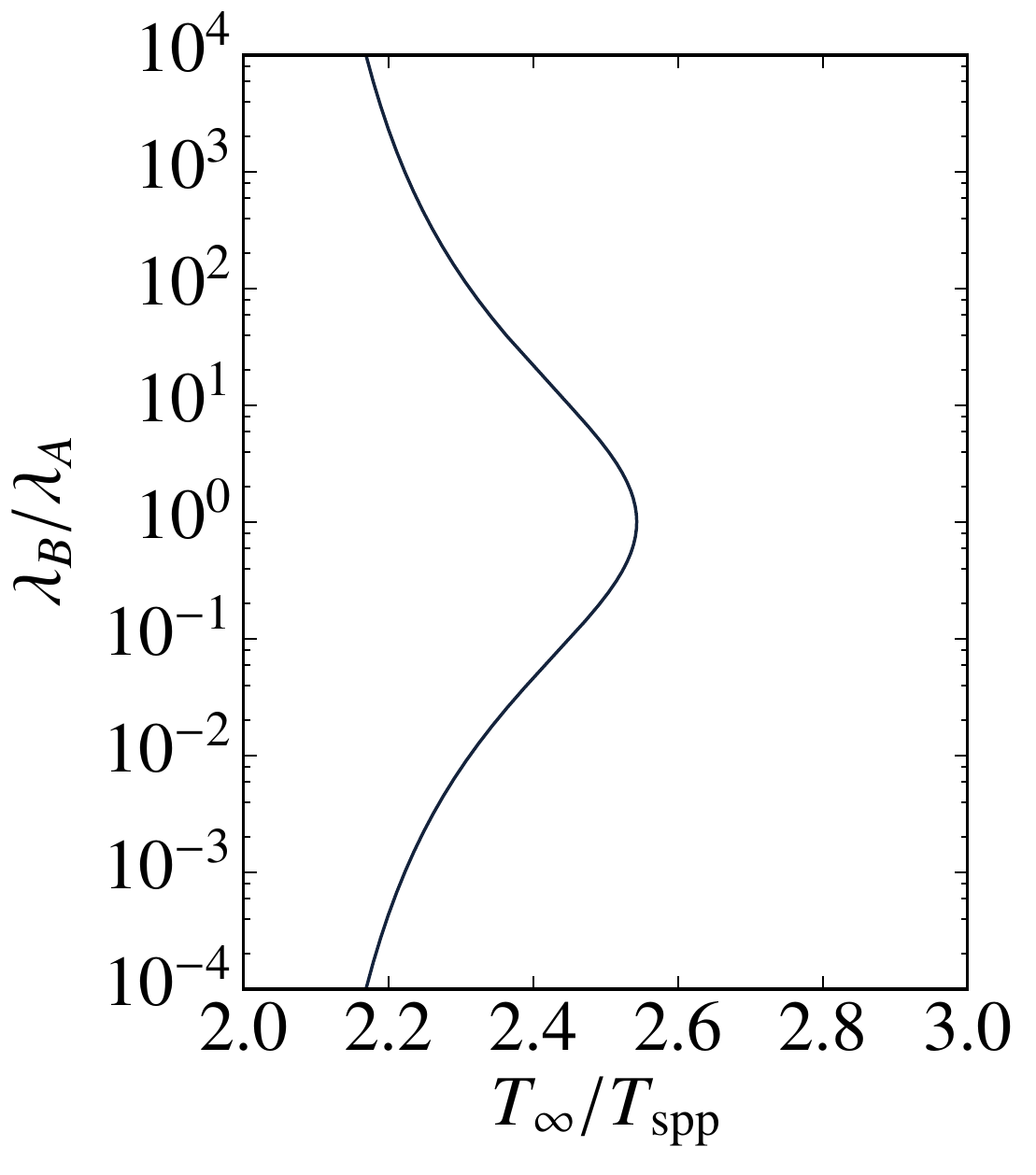}
        \label{fig:full_phase_diagram:ratio}
    }%
    \caption{(a) Entanglement phase diagram for the two-dimensional toric code, as predicted by the negativity. As temperature is increased, defects are thermally excited and the negativity is degraded until, at $T_\text{spp}$, the SPP negativity vanishes identically. However, entanglement between subsystems that share a boundary with $N_\partial > 1$ is more robust, and able to survive above $T_\text{spp}$. When $T_\infty$ is reached, the negativity vanishes identically even for strip partitions that span the entire system. (b) Ratio $T_\infty / T_\text{spp}$ of the two limiting temperatures.}
    \label{fig:full_phase_diagram}
\end{figure}

Writing the sudden death condition in terms of the number density of each type of excitation, and defining for convenience $\varDelta_X \equiv n_h(2\lambda_X) - n_\text{F}(2\lambda_X) > 0$, one obtains 
\begin{equation}
    \frac{n_\text{F}(2\lambda_A) n_\text{F}(2\lambda_B) }{(\varDelta_A \varDelta_B)^{1/2}}  
		= 
		\cos\left( \frac{\pi}{2N_\partial} \right)
		\, . 
    \label{eqn:sudden_death_condition_physical_larger_boundary}
\end{equation}
That is, the negativity vanishes when the geometric mean of the number densities of both types of excitations, each weighted by $\varDelta_X^{-1/2}$, equals or exceeds an $\mc{O}(1)$ number (which tends towards unity in the limit of a long boundary). The similarity between Eq.~\eqref{eqn:sudden_death_condition_physical} and Eq.~\eqref{eqn:sudden_death_condition_physical_larger_boundary} is apparent. Arguably the most important conclusion we can draw from Eq.~\eqref{eqn:sudden_death_condition_physical_larger_boundary} is that the critical temperature $T_c$ remains finite in the limit of large boundaries $N_\partial \to \infty$: indeed $T_c \sim \lambda_B$, as in the SPP case.
For a system-spanning strip, the limiting behaviour of $T_c$ is
\begin{equation}
    \frac{T_c}{\lambda_A} \simeq \dfrac{4\frac{\lambda_B}{\lambda_A}}{W\left(\frac{\lambda_B}{\lambda_A}\right)}
    \quad \text{ for } \quad \lambda_B \gg \lambda_A, \; N_\partial \to \infty
        \, ,
\end{equation}
which is to be contrasted with Eq.~\eqref{eqn:SPP_sudden_death_limit}. 

Our main findings are embodied by the entanglement ``phase diagram'' in Fig.~\ref{fig:full_phase_diagram}. The entanglement between subsystems that share an extended boundary is always more robust to the effects of thermal fluctuations than the SPP system, but the two sudden death temperatures only differ by a factor of approximately two, as shown in Fig.~\ref{fig:full_phase_diagram:ratio}. From these results, it seems reasonable to expect that for all finite $\lambda_B / \lambda_A$, there exists a temperature $T_\infty$ above which the negativity vanishes for any tripartition of the lattice. We expect this temperature to be controlled by the largest gap to elementary excitations in the system, up to a prefactor of $\mc{O}(1)$.

In Sec.~\ref{sec:larger_boundary_calculation} we verify that these results remain qualitatively unchanged for a \emph{contractible} subsystem, and moreover that both contractible and system-spanning subsystems give rise to the same sudden death temperature $T_\infty$ in the limit of infinite boundaries $N_\partial \to \infty$.


\subsection{$\lambda_B \to \infty$ limit}
\label{sec:infinite_lambda_B}

When one of the two coupling constants in the model is infinite, $\lambda_B \to \infty$, the formation of magnetic vortices is rendered energetically impossible for any finite temperature. Conversely, electric charges are present in the system.
This hard constraint means that the subspace of thermally accessible states is substantially smaller than the full Hilbert space, and the calculation is greatly simplified. For the partition scheme in Fig.~\ref{fig:partitions:tripartite}, we arrive at the expression (see Sec.~\ref{sec:infinite_lambda_B_calculation} for details)
\begin{equation}
    \mathrm{e}^{\mc{E}(T)} = \frac{\cosh\left(\frac12 (N - N_\partial) K_A \right)}{\cosh\left(\frac12 N K_A\right)} \left[ 2\cosh\left(\frac{K_A}{2}\right) \right]^{N_\partial}
		\, ,
    \label{eqn:tripartite_negativity_final}
\end{equation}
where the temperature dependence is controlled by the parameter $K_A \equiv - \ln\tanh\beta\lambda_A$. In this limit, the negativity does not exhibit any discontinuities. When evaluated for $N_\partial=1$, this expression correctly agrees with the SPP result~\eqref{eqn:single_star_negativity} in the appropriate limit $\lambda_B \to \infty$.

If we take the thermodynamic limit $N \to \infty$ while keeping the length of the boundary $N_\partial$ finite, this expression simplifies further, and its extensivity in the number of boundary stars becomes apparent.
Introducing the density of nondefective stars $n_h$, the negativity becomes
\begin{equation}
    \mc{E}(T) = N_\partial \ln\left( 2n_h \right)
        \, .
    \label{eqn:tripartite_negativity_final_thermodynamic}
\end{equation}
For \emph{all} temperatures, the negativity for $N \to \infty$ satisfies an exact area law.

At zero temperature, the density of excitations is zero, or $n_h=1$, and the negativity assumes the value $\mc{E}(0) = N_\partial \ln 2$, in agreement with the results of Ref.~\onlinecite{Lee2013}. Since the subsystem $\mc{A}$ is now considered contractible, the topological contribution seen in Sec.~\ref{sec:larger_boundary} is removed. If the subsystems $\mc{A}_1$ and $\mc{A}_2$ do not share a boundary then the negativity vanishes identically for all temperatures.

The $1/T$ decay of the negativity above $T \sim \lambda_A$ that was observed in the SPP is no longer cut off at high temperatures. Sending $\lambda_B \to \infty$ prevents the thermal excitation of magnetic defects, thereby locking in the zero-temperature magnetic loop structure. As a result, thermal fluctuations are never sufficient to rid the system of all quantum entanglement.

Since the ground states are independent of the system's parameters, so is $\mc{E}(0)$. This peculiarity of the toric code means that entanglement is able to survive up to arbitrarily large temperatures, suffering only a slow $1/T$ demise. In other models this is not always possible as divergence of the sudden death temperature is often accompanied by a vanishing zero-temperature negativity. This is the case in the 1D transverse field Ising model, for example, which we discuss in Appendix~\ref{sec:ising_model}.


\section{calculations}
\label{sec:calculations}

We begin this section with some general considerations about the calculation of the entanglement negativity $\mc{E}(T)$ of the toric code. We will then consider specific cases that allow us to streamline the calculations and shape the final results in the relatively simple and insightful formulae discussed in Sec.~\ref{sec:main_results}.

We consider a system of $2N$ spins in thermodynamic equilibrium with a thermal reservoir at temperature $T = 1/\beta$ (we set the Boltzmann constant $k_\text{B}=1$ throughout). The system is therefore described by the canonical density matrix $\rho = \mathrm{e}^{- \beta H} / Z$, where $H$ is the Hamiltonian and $Z = \Tr \mathrm{e}^{- \beta H}$ is the partition function (to ensure proper normalisation, $\Tr \rho = 1$). The density matrix can be written in terms of its matrix elements with respect to complete bases $\{ \ket*{\alpha} \}$ and $\{ \ket*{\gamma} \}$ as
\begin{equation}
    \rho = \frac{1}{Z} \sum_{\alpha,\gamma} \mel*{\alpha}{\mathrm{e}^{-\beta H}}{\gamma} \ket*{\alpha} \bra*{\gamma} 
		\, . 
    \label{eqn:canonical_density_matrix}
\end{equation}

Following the work in Refs.~\onlinecite{Castelnovo2013, Castelnovo2008-Feb, Castelnovo2007-174416, Castelnovo2007, Hamma2005}, we choose the tensor product basis of eigenstates of the operator $\otimes_i \sigma_i^z$. The operators $B_p \equiv \prod_{i \in p} \sigma_i^z$ are diagonal in this basis: $P\ket*{\alpha} = M(\alpha)\ket*{\alpha}$, where one can interpret the eigenvalue $M(\alpha)$ of $P=\sum_p B_p$ as a ``plaquette magnetisation'' equal to the sum of all local magnetisations $M_p \equiv \mel*{\alpha}{B_p}{\alpha}$.
We then introduce the group $\mc{G}$ generated by products of star operators $A_s$. The group elements $g \in \mc{G}$ must be defined modulo the identity for $g^{-1}$ to be unique, since $\prod_s A_s = \mathds{1}$. Only states that differ by the action of a group element $g \in \mc{G}$ give nonvanishing matrix elements in~\eqref{eqn:canonical_density_matrix}, which allows us to write $\ket*{\gamma} = g\ket*{\alpha}$. Hence~\cite{Castelnovo2013, Castelnovo2008-Feb, Castelnovo2007-174416, Castelnovo2007, Hamma2005},
\begin{equation}
    \rho = \frac{1}{Z} \sum_{\alpha} \sum_{g \in \mc{G}} \mathrm{e}^{\beta\lambda_BM(\alpha)} \mel*{\alpha}{\mathrm{e}^{\beta S} g}{\alpha} \ket*{\alpha} \bra*{\alpha} g 
		\, .
    \label{eqn:density_matrix_star_plaquette_split}
\end{equation}
Following similar considerations, the partition function $Z$ can be written as 
\begin{align}
    Z &= (\cosh\beta\lambda_A)^N [1 + (\tanh\beta\lambda_A)^N] \cdot \sum_\alpha \mathrm{e}^{\beta\lambda_B M(\alpha)} 
		\\
    &\equiv Z_A \cdot Z_B 
		\, . 
\end{align}
Notice that the system is symmetric upon exchanging star and plaquette operators, and $\lambda_A \leftrightarrow \lambda_B$. Therefore, it would be completely equivalent to choose the tensor product basis of the operator $\otimes_i \sigma_i^x$, and so we must have $Z_A(\beta\lambda) \propto Z_B(\beta\lambda)$.

To evaluate the matrix elements in~\eqref{eqn:density_matrix_star_plaquette_split}, it is helpful to expand the exponential in terms of its constituent star operators. Remembering that $S=\sum_s A_s$ and that $ A_s^2 = \mathds{1}$,
\begin{equation}
	\mathrm{e}^{\beta \lambda_A S} = \prod_{s} \left[ \cosh(\beta\lambda_A) + \sinh(\beta\lambda_A) A_s \right] 
	\, , 
\end{equation}
and we thus obtain 
\begin{equation}
	\frac{1}{Z_A} \mathrm{e}^{\beta \lambda_A S} = \sum_{\tilde{g} \in \mc{G}} \eta_T(\tilde{g}) \tilde{g} 
	\, .
\end{equation}
Here we introduced for convenience of notation the weighting factor 
\begin{equation}
    \eta_T (g) = \frac{\mathrm{e}^{-K_A \mathscr{n}(g)} + \mathrm{e}^{-K_A \left[ N - \mathscr{n}(g) \right]}}{1 + \mathrm{e}^{-K_A N}} 
		\, , 
    \label{eqn:eta_T_definition}
\end{equation}
where $K_A \equiv -\ln\tanh\beta\lambda_A$ and $\mathscr{n}(g)$ is the number of star operators $A_s$ that appear in the decomposition of the group element $g$. Note that $\eta_T(g)$ is invariant under $\mathscr{n}(g) \to N - \mathscr{n}(g)$, and therefore the ambiguity in the definition of the group elements of $\mc{G}$ modulo the identity is immaterial~\cite{Castelnovo2013, Castelnovo2008-Feb, Castelnovo2007-174416, Castelnovo2007, Hamma2005}.
We then arrive at the expression for the density matrix 
\begin{equation}
    \rho = \frac{1}{Z_B}\sum_\alpha \sum_{g \in \mc{G}} \mathrm{e}^{\beta\lambda_B M(\alpha)} \eta_T(g) \ket*{\alpha} \bra*{\alpha} g 
		\, .
    \label{eqn:density_matrix_g_form}
\end{equation}

If we are interested only in a subsystem $\mc{A}$ of the total system $\mc{S}$, then we should trace out the complementary subsystem $\mc{B}$ to form the reduced density matrix $\rho_\mc{A} = \Tr_\mc{B} \rho$. To achieve this, we decompose $\ket*{\alpha} = \ket*{\alpha_\mc{A}}\otimes \ket*{\alpha_\mc{B}}$ and $g = g_\mc{A} \otimes g_\mc{B}$,
\begin{multline}
    \rho_\mc{A} = \frac{1}{Z_B}\sum_\alpha \sum_{g \in \mc{G}} \mathrm{e}^{\beta\lambda_B M(\alpha)} \eta_T(g)  
		\; \times 
		\\
		\mel*{\alpha_\mc{B}}{g_\mc{B}}{\alpha_\mc{B}} \ket*{\alpha_\mc{A}} \bra*{\alpha_\mc{A}} g_\mc{A} 
		\, .
    \label{eqn:reduced_density_matrix}
\end{multline}
Now, the matrix element equals unity if $g$ acts trivially on the $\mc{B}$ subsystem, and zero otherwise. The resulting constraint on $g$ is implemented by restricting the summation to group elements belonging to the subgroup $\{ g \in \mc{G} \;|\; g_\mc{B}=\mathds{I}_\mc{B} \} \equiv \mc{G}_\mc{A}  \subset \mc{G}$.

Finally, we recall that $\mc{A}$ is further partitioned into two subsystems ($\mc{A} = \mc{A}_1 \cup \mc{A}_2$), and that we want to take the partial transpose over one of them (say, $\mc{A}_2$).
This is effected by splitting up the states $\ket*{\alpha_\mc{A}} = \ket*{\alpha_{\mc{A}_1}} \otimes \ket*{\alpha_{\mc{A}_2}}$, and similarly for the group elements $g_\mc{A} = g_{\mc{A}_1} \otimes g_{\mc{A}_2}$
\begin{multline}
    \rho_\mc{A}^{T_2} =  \frac{1}{Z_B} \sum_{\alpha} \sum_{g \in \mc{G}_\mc{A}}
        \mathrm{e}^{\beta\lambda_B M(\alpha)} \eta_T(g)
        \; \times
        \\
        \left( \ket*{\alpha_{\mc{A}_1}} \bra*{\alpha_{\mc{A}_1}} g_{\mc{A}_1} \right) \otimes \left( g_{\mc{A}_2} \ket*{\alpha_{\mc{A}_2}} \bra*{\alpha_{\mc{A}_2}} \right) 
		\, .
		\label{eqn:reduced_density_matrix_T2}
\end{multline}
We can then apply the replica approach~\eqref{eqn:replica_approach_expression} by taking the trace of the $n$th power of this operator,
\begin{widetext}
\begin{equation}
\begin{aligned}
    \Tr\big(\rho_\mc{A}^{T_2}\big)^n = \frac{1}{Z_B^n} \sum_{\alpha_{1}, \ldots, \alpha_{n}}& \sum_{g_1, \ldots, g_n \in \mc{G}_\mc{A}} \left( \prod_{\ell = 1}^n  \mathrm{e}^{\beta\lambda_B M(\alpha_{\ell})} \eta_T(g_\ell) \right)  \\ &\mel*{\alpha_{1\mc{A}_1}}{g_{1\mc{A}_1}}{\alpha_{2\mc{A}_1}} \cdots \mel*{\alpha_{(n-1)\mc{A}_1}}{g_{(n-1)\mc{A}_1}}{\alpha_{n\mc{A}_1}}  \mel*{\alpha_{n\mc{A}_1}}{g_{n\mc{A}_1}}{\alpha_{1\mc{A}_1}}  \\
    &\mel*{\alpha_{1\mc{A}_2}}{g_{2\mc{A}_2}}{\alpha_{2\mc{A}_2}} \cdots \mel*{\alpha_{(n-1)\mc{A}_2}}{g_{n\mc{A}_2}}{\alpha_{n\mc{A}_2}}  \mel*{\alpha_{n\mc{A}_2}}{g_{1\mc{A}_2}}{\alpha_{1\mc{A}_2}} 
		\, .
\end{aligned}
\label{eqn:trace_full}
\end{equation}
\end{widetext}

It is helpful to notice that the subgroup $\mc{G}_\mc{A}$ can in general be decomposed as $\mc{G}_\mc{A} = \mc{G}_{\mc{A}_1} \mc{G}_{\mc{A}_2} \mc{G}_{\mc{A}_1\mc{A}_2}$. The subgroups $\mc{G}_{\mc{A}_i}$ are defined as $\mc{G}_{\mc{A}_i} \equiv \{ g \in \mc{G}_\mc{A} \;|\; g_{\bar{\mc{A}}_i} = \mathds{I}_{\bar{\mc{A}}_i} \}$ with $\bar{\mc{A}}_1 = \mc{A}_2$ and vice versa, while $\mc{G}_{\mc{A}_1\mc{A}_2}\equiv \mc{G}_\mc{A} / (\mc{G}_{\mc{A}_1} \mc{G}_{\mc{A}_2})$ is the quotient group. Any element $g \in \mc{G}_\mc{A}$ can therefore be uniquely decomposed into the product of three group elements: one that acts exclusively on $\mc{A}_1$, one exclusively on $\mc{A}_2$, and one that acts simultaneously (and exclusively) on $\mc{A}_1$ and $\mc{A}_2$. Namely, $g = \bar{g}\bar{\bar{g}}\theta$, with $\bar{g} \in \mc{G}_{\mc{A}_1}$, $\bar{\bar{g}} \in \mc{G}_{\mc{A}_2}$, and $\theta \in \mc{G}_{\mc{A}_1\mc{A}_2}$~\cite{Castelnovo2013}.

Using this decomposition, we make the following relabelling of the states $\ket*{\alpha_k}$, for all $k>1$:
\begin{equation}
    \ket*{\alpha_k^\prime} = \left( \prod_{\ell=1}^{k-1} \bar{g}_\ell \right)\left( \prod_{\ell=2}^k \bar{\bar{g}}_\ell \right) \ket*{\alpha_k} \to \ket*{\alpha_k}
		\, .
\end{equation}
This removes all dependence on $\bar{g}_\ell$ and $\bar{\bar{g}}_\ell$ from the matrix elements, and we obtain a more compact expression:
\begin{widetext}
\begin{equation}
    \Tr\big(\rho_\mc{A}^{T_2}\big)^n = \frac{1}{Z_B^n}\sum_{\alpha_{1}, \ldots, \alpha_{n}} \sum_{\substack{\bar{g}_1, \ldots, \bar{g}_n \\ \in \mc{G}_{\mc{A}_1}}} \sum_{\substack{\bar{\bar{g}}_1, \ldots, \bar{\bar{g}}_n \\ \in \mc{G}_{\mc{A}_2}}} \sum_{\substack{\theta_1, \ldots, \theta_n \\ \in \mc{G}_{\mc{A}_1\mc{A}_2}}} \mel*{0}{\prod_{\ell=1}^n \bar{g}_\ell \bar{\bar{g}}_\ell }{0}
    \left( \prod_{\ell = 1}^n  \mathrm{e}^{\beta\lambda_B M(\alpha_{\ell})}  \eta_T(\bar{g}_\ell\bar{\bar{g}}_\ell \theta_\ell) \mel*{\alpha_{\ell\mc{A}}}{\theta_{\ell\mc{A}_1} \otimes \theta_{(\ell  + 1)\mc{A}_2} }{\alpha_{(\ell+1)\mc{A}}} \right)
		\, ,
    \label{eqn:trace_full_after_redefinition}
\end{equation}
\end{widetext}
where $\theta_{n+1}\equiv\theta_1$ and similarly $\alpha_{n+1}\equiv\alpha_1$. Notice that the matrix elements in Eq.~\eqref{eqn:trace_full_after_redefinition} impose constraints on which terms $\bar{g}_\ell$ and $\bar{\bar{g}}_\ell$ in the summation give a nonvanishing contribution. Furthermore, they implicitly impose the constraint $\prod_\ell \theta_\ell = \mathds{I}$. This is because, given $\alpha_1$ (say), it takes only $n-1$ of the $n$ matrix elements to uniquely determine all other $\alpha_\ell$; the final matrix element then evaluates to $\mel*{\alpha_1}{\prod_\ell \theta_\ell}{\alpha_1}$, which indeed is nonvanishing only if the aforementioned constraint is satisfied. 
\begin{figure}[b]
    \centering
    \subfloat[]{%
        \includegraphics[width=0.4225\linewidth]{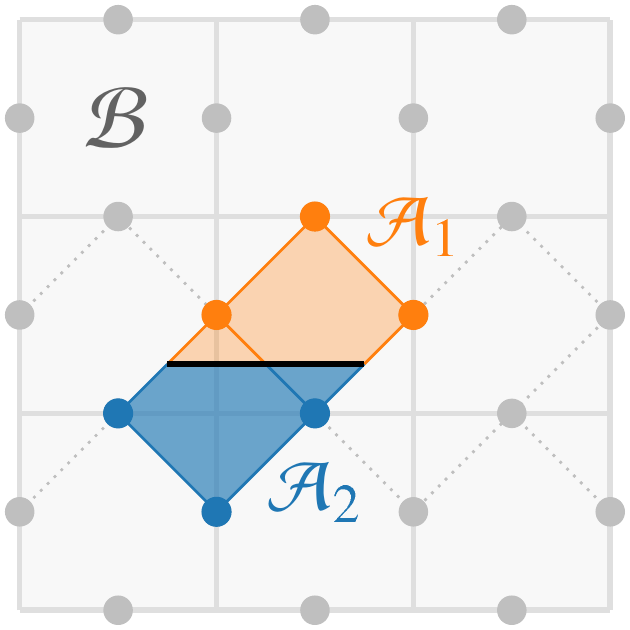}
        \label{fig:boundary_length_lattice:spp}
    }%
    \hfill
    \subfloat[]{%
        \includegraphics[width=0.555\linewidth]{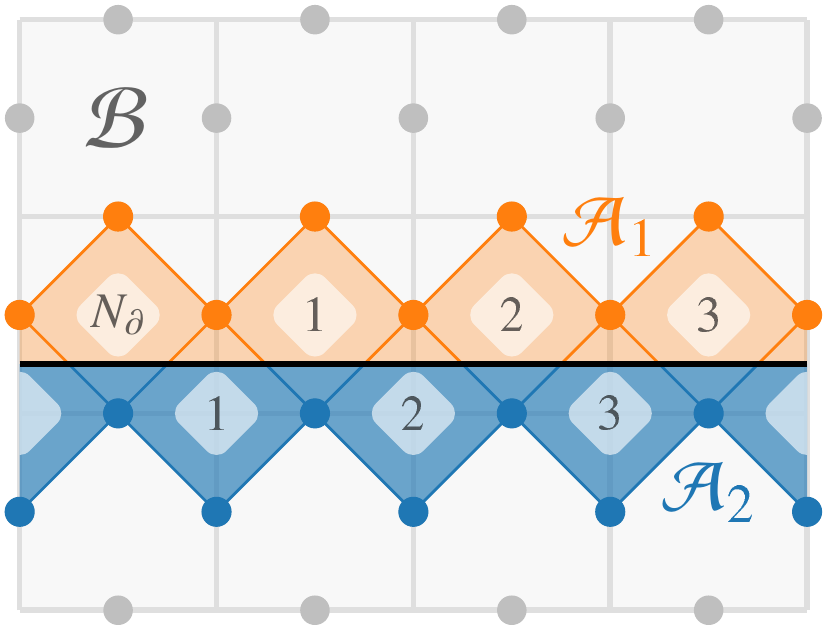}
        \label{fig:boundary_length_lattice:extended}
    }%
    \caption{Illustration of (a) a star plaquette pair and (b) a strip partition chosen for the calculation of $\mc{E}$ when the boundary length $N_\partial > 1$. In both cases, the subsystem $\mc{A}=\mc{A}_1 \cup \mc{A}_2$ is symmetric under interchange of stars and plaquettes, and in (b) it spans the torus in one direction.}
    \label{fig:boundary_length_lattice}
\end{figure}

Equation~\eqref{eqn:trace_full_after_redefinition} represents the general form of $\Tr\big(\rho_\mc{A}^{T_2}\big)^n$ for a tripartition of the system, and is the starting point of all replica method calculations discussed later.


\subsection{Star plaquette pair}
\label{sec:single_star_calculation}

It is convenient to consider first the case of the smallest nontrivial subsystem: a single star plaquette pair, shown in Fig.~\ref{fig:boundary_length_lattice:spp}.
Notice that subsystem $\mc{A}$ is symmetric under interchange of stars with plaquettes, which has the important implication that $\mc{E}$ must be symmetric upon exchanging $\lambda_A \leftrightarrow \lambda_B$.
This property will be used at the end of this section: given the dependence of $\mc{E}$ on $\lambda_A$, say, one can then deduce the dependence on $\lambda_B$ from the relation $\mc{E}(\lambda_A, \lambda_B) = \mc{E}(\lambda_B, \lambda_A)$.

Rather than proceeding via the replica method, it is more convenient in this case to diagonalise the partial transpose $\rho_\mc{A}^{T_2}$ of the reduced density matrix explicitly. We start with the general expression for the partially transposed reduced density matrix~\eqref{eqn:reduced_density_matrix_T2} derived earlier:
\begin{multline}
    \rho_\mc{A}^{T_2} =  \sum_{\alpha_\mc{A}} \sum_{g \in \mc{G}_\mc{A}} \mathrm{e}^{\beta\lambda_B M_\partial (\alpha_\mc{A})} \xi_T(M_\partial) \eta_T(g) \\
    \left( \ket*{\alpha_{\mc{A}_1}} \bra*{\alpha_{\mc{A}_1}} g_{\mc{A}_1} \right) \otimes \left( g_{\mc{A}_2} \ket*{\alpha_{\mc{A}_2}} \bra*{\alpha_{\mc{A}_2}} \right)
        \, .
    \label{eqn:reduced_density_matrix_T2_xi}
\end{multline}
The sum over spins $i \in \mc{B}$ has been absorbed into the factors $\xi_T(M_\partial) \equiv \frac{1}{Z_B} \sum_{\alpha_\mc{B}} \mathrm{e}^{\beta \lambda_B M_{\bar{\partial}}(\alpha)}$ by splitting the magnetisation into a boundary component $M_\partial \equiv \ev*{B_{p \in \partial}}$ and its complement $M_{\bar{\partial}} = \sum_{p\notin \partial} \ev*{B_p}$ so that $M(\alpha) = M_\partial(\alpha_\mc{A}) + M_{\bar{\partial}}(\alpha)$. Although at first sight it may appear that $\xi_T$ could assume a different value for each spin configuration $\alpha_\mc{A}$, it turns out to depend only on the magnetisation $M_\partial$ of the boundary plaquette, as the notation indicates. For brevity we ask the reader to take this result for granted until Sec.~\ref{sec:larger_boundary_calculation}, where we provide a proof.

The subsystem contains only one star which acts simultaneously on $\mc{A}_1$ and $\mc{A}_2$. Therefore the group $\mc{G}_\mc{A}$ coincides with $\mc{G}_{\mc{A}_1\mc{A}_2}$ and contains only one nontrivial element: the boundary star operator $A_s$. Decomposing it into components that act separately on subsystems $\mc{A}_1$ and $\mc{A}_2$, $A_s = A_s^{(1)} \otimes A_s^{(2)}$, Eq.~\eqref{eqn:reduced_density_matrix_T2_xi} can be written as
\begin{multline}
    \rho_\mc{A}^{T_2} = \sum_{\alpha_\mc{A}} \mathrm{e}^{\beta\lambda_B M_\partial(\alpha_\mc{A}) } \xi_T(M_\partial) \Big( \ket*{\alpha_\mc{A}} \bra*{\alpha_\mc{A}} + \\ \eta_A \ket*{\alpha_{\mc{A}_1} \otimes A_s^{(2)} \alpha_{\mc{A}_2}} \bra*{A_s^{(1)} \alpha_{ \mc{A}_1} \otimes  \alpha_{\mc{A}_2}} \Big)
		\, ,
	\label{eqn:SPP_reduced_density_matrix_T2_xi}
\end{multline}
where $\eta_A$ was defined in Eq.~\eqref{eqn:eta_redefined}. Relabelling $\alpha_{\mc{A}_1} \otimes A_s^{(2)} \alpha_{\mc{A}_2} = \alpha_\mc{A}^\prime$ in the second term, we obtain
\begin{equation}
    \ket*{\alpha_{\mc{A}_1} \otimes A_s^{(2)} \alpha_{\mc{A}_2}} \bra*{A_s^{(1)} \alpha_{ \mc{A}_1} \otimes  \alpha_{\mc{A}_2}}
		=
		\ket*{\alpha^\prime_{\mc{A}} } \bra*{ A_s \alpha^\prime_{\mc{A}} }
		\, .
\end{equation}
We can then change the mute index $\alpha^\prime_{\mc{A}} \to \alpha_\mc{A}$ and denote $\ket*{\bar{\alpha}_\mc{A}} = A_s \ket*{\alpha_\mc{A}}$ to obtain
\begin{multline}
    \rho_\mc{A}^{T_2} = \sum_{\alpha_\mc{A}} \mathrm{e}^{\beta\lambda_B M_\partial(\alpha_\mc{A}) } \xi_T(M_\partial) \ket*{\alpha_\mc{A}} \bra*{\alpha_\mc{A}} + \\ \eta_A \mathrm{e}^{-\beta\lambda_B M_\partial(\alpha_\mc{A}) } \xi_T(-M_\partial) \ket*{\alpha_{\mc{A}} } \bra*{ \bar{\alpha}_{\mc{A}} }
		\, .
    \label{eqn:partial_transposed_explicit}
\end{multline}
Each state $\ket*{\alpha_\mc{A}}$ of the subsystem $\mc{A}$ is coupled only to itself and its counterpart with the spins $i \in s$ flipped. The expression~\eqref{eqn:partial_transposed_explicit} implies that $\rho_\mc{A}^{T_2}$ can be written as a block diagonal matrix with respect to the eigenstates of $\otimes_{i \in \mc{A}} \sigma_i^z$, with the $2\times 2$ matrices
\begin{equation}
    \rho_\pm = \begin{pmatrix}
        \Xi^\pm & \eta_A \Xi^\mp \\
        \eta_A \Xi^\mp & \Xi^\pm
    \end{pmatrix}
    \label{eqn:lambda_B_finite_block}
\end{equation}
along the diagonal (note that these are not density matrices). For convenience we introduced the shorthand $\Xi^\pm$ to represent $\xi_T(M_\partial) \mathrm{e}^{\beta\lambda_B M_\partial}$ evaluated with boundary magnetisation $M_\partial = \pm 1$. Since the number of spin configurations $\alpha_\mc{A}$ with $M_\partial = +1$ equals the number with $M_\partial = -1$, there exist an equal number of $\rho_+$ and $\rho_-$ blocks along the diagonal of $\rho_{\mc{A}}^{T_2}$. Therefore the normalisation condition $\Tr\rho_\mc{A}^{T_2} = 1$ gives $\Xi^+ + \Xi^- = 2^{1-\mc{N}_\mc{A}}$, where $\mc{N}_\mc{A} \equiv 6$ equals the number of spins $i \in \mc{A}$.

After calculating the eigenvalues of $\rho_\pm$, invoking the symmetry $\mc{E}(\lambda_A, \lambda_B) = \mc{E}(\lambda_A, \lambda_B)$ allows us to conjecture that $\Xi^+ - \Xi^- = 2^{1-\mc{N}_\mc{A}}\eta_B$. This implies that $\Xi^\pm = (1 \pm \eta_B) / 2^{\mc{N}_\mc{A}}$, and
\begin{equation}
    \mathrm{e}^{\mc{E}(T)} = 1 + \tfrac12 \max(\eta_A + \eta_B + \eta_A\eta_B - 1, 0)
    \, .
    \label{eqn:SPP_negativity_calculation_final}
\end{equation}

Alternatively, one can follow the (lengthier) calculations in  Sec.~\ref{sec:larger_boundary_calculation} to evaluate $\Xi^\pm$ explicitly; we find indeed that the latter approach reproduces~\eqref{eqn:SPP_negativity_calculation_final} exactly, without relying on the symmetry argument.

That the negativity for the SPP is both a necessary and sufficient condition for separability follows almost trivially from the fact that $\mc{E} = 0$ is a sufficient condition for separability when $\mc{H} = \mathbb{C}^2 \otimes \mathbb{C}^2$~\cite{Horodecki1996}. In a similar manner to~\eqref{eqn:SPP_reduced_density_matrix_T2_xi}, the reduced density matrix of the SPP is
\begin{equation}
    \rho_\mc{A} =  \sum_{\alpha_\mc{A}} \mathrm{e}^{\beta\lambda_B M_\partial (\alpha_\mc{A})} \xi_T(M_\partial)
    \Big( \ket*{\alpha_\mc{A}} \bra*{\alpha_\mc{A}} + \eta_A \ket*{\alpha_\mc{A}} \bra*{\alpha_\mc{A}}A_s \Big)
        \, .
    \label{eqn:SPP_reduced_density_matrix_expanded}
\end{equation}
We then partition the Hilbert space, $\mc{H}_\mc{A}$, of the SPP into a direct sum of ($16 \equiv k$) 4-dimensional subspaces $\mc{H}_\mc{A} = \mc{H}_1 \oplus \mc{H}_2 \oplus \ldots \oplus \mc{H}_k$. Each $\mc{H}_i$ is taken to be of the form
\begin{equation}
    \Span \left\{ \ket*{\alpha_\mc{A}}, A_s^{(1)} \ket*{\alpha_\mc{A}}, A_s^{(2)} \ket*{\alpha_\mc{A}}, A_s \ket*{\alpha_\mc{A}} \right\}
        \, ,
    \label{eqn:SPP_subspace}
\end{equation}
for some state $\alpha$. Inspection of \eqref{eqn:SPP_reduced_density_matrix_expanded} shows that one can also write the density matrix as a direct sum
\begin{equation}
    \rho_\mc{A} = \frac{1}{k} \bigoplus_{i=1}^k \varrho
        \, ,
    \label{eqn:SPP_density_matrix_direct_sum}
\end{equation}
where we have introduced the density matrix
\begin{equation}
    \varrho = k
    \begin{pmatrix}
        \Xi^+ & 0 & 0 & \eta_A \Xi^+ \\
        0 & \Xi^- & \eta_A \Xi^- & 0 \\
        0 & \eta_A \Xi^- & \Xi^- & 0 \\
        \eta_A \Xi^+ & 0 & 0 & \Xi^+
    \end{pmatrix}
        \, .
\end{equation}
By partitioning $\mc{H}_\mc{A}$ according to \eqref{eqn:SPP_subspace}, partial transposition acts on each $\varrho$ separately:
\begin{equation}
    \rho_\mc{A}^{T_2} = \frac{1}{k} \bigoplus_{i=1}^k \varrho^{T_2}
        \, ,
\end{equation}
and consequently $\mc{E}(\rho_\mc{A}) = \mc{E}(\varrho)$. We are now able to make use of the result that the negativity $\mc{E}$ is both a necessary and sufficient condition for separability when $\dim{\mc{H}} = 2 \times 2$~\cite{Horodecki1996}. If $\mc{E}(\varrho) = 0$, then the state $\varrho$ is separable and, by definition, can be written $\varrho = \sum_j p_j (\varrho_j^{\mc{A}_1} \otimes \varrho_j^{\mc{A}_2})$ with $\sum_j p_j = 1$. Substituting into \eqref{eqn:SPP_density_matrix_direct_sum}, after some manipulation, one arrives at
\begin{align}
    \rho_\mc{A} &= \sum_j p_j \left( \frac{1}{\sqrt{k}} \bigoplus_{i=1}^{\sqrt{k}} \varrho_j^{\mc{A}_1} \right) \otimes \left( \frac{1}{\sqrt{k}} \bigoplus_{i=1}^{\sqrt{k}} \varrho_j^{\mc{A}_2} \right) \nonumber \\
        &\equiv \sum_j p_j \rho_j^{\mc{A}_1} \otimes \rho_j^{\mc{A}_2}
    \, .
\end{align}
That is, the state $\rho_\mc{A}$ is separable if its negativity $\mc{E}(\rho_\mc{A})$ vanishes. We remark that this is a consequence of the very special form of the reduced density matrix described by Eq.~\eqref{eqn:SPP_density_matrix_direct_sum}.


\subsection{Extended boundary}
\label{sec:larger_boundary_calculation}

Let us now consider the more general case where the $\mc{A}_1$--$\mc{A}_2$ boundary spans the entire system, as illustrated schematically in Fig.~\ref{fig:boundary_length_lattice:extended}, leading to the result~\eqref{eqn:negativity_larger_boundary} discussed in Sec.~\ref{sec:larger_boundary}. Both $\mc{A}_1$ and $\mc{A}_2$ span the torus in one direction and share one edge. The subsystem $\mc{A}\equiv \mc{A}_1 \cup \mc{A}_2$ consists only of boundary stars ($A_{s \in \partial}$) and boundary plaquettes ($B_{p\in \partial}$) which act simultaneously (and exclusively) on both the $\mc{A}_1$ and $\mc{A}_2$ subsystems. In addition, $\mc{A}$ is symmetric under interchange of stars and plaquettes so, as for the SPP system, the final result must obey the symmetry $\mc{E}(\lambda_A, \lambda_B) = \mc{E}(\lambda_B, \lambda_A)$.

The calculation starts from the most general expression~\eqref{eqn:trace_full_after_redefinition} for $\Tr\big(\rho_\mc{A}^{T_2}\big)^n$. Since $\mc{A}$ contains only boundary stars, the groups $\mc{G}_{\mc{A}_1}$ and $\mc{G}_{\mc{A}_2}$ become trivial ($\{ \mathds{I} \}$), removing the summations over $\bar{g}_\ell$ and $\bar{\bar{g}}_\ell$. In addition, the magnetisation can be decomposed into a boundary component $M_\partial$, equal to the sum of the magnetisations of all boundary plaquettes straddling $\mc{A}_1$ and $\mc{A}_2$, and its complement: $M(\alpha) = M_\partial(\alpha_\mc{A}) + M_{\bar{\partial}}(\alpha)$.
Note that the boundary magnetisation $M_\partial$ depends only on spins contained within subsystem $\mc{A}$ while $M_{\bar{\partial}}$ depends on all spins.
Similarly to the SPP, tracing over subsystem $\mc{B}$ is then entirely contained within a factor $\xi_T(\alpha_\mc{A}) \equiv \frac{1}{Z_B} \sum_{\alpha_\mc{B}} \mathrm{e}^{\beta \lambda_B M_{\bar{\partial}}(\alpha)}$, whose explicit calculation is left until later. After making these simplifications, \eqref{eqn:trace_full_after_redefinition} becomes
\begin{multline}
    \Tr\big(\rho_\mc{A}^{T_2}\big)^n = \sum_{\alpha_{1\mc{A}}, \ldots, \alpha_{n\mc{A}}} \sum_{\substack{\theta_1, \ldots, \theta_n \\ \in \mc{G}_{\mc{A}_1\mc{A}_2}}} \mel*{0}{\prod_{\ell=1}^n \theta_\ell }{0} \mathrm{e}^{\beta\lambda_B \sum_\ell M_\partial(\alpha_{\ell\mc{A}})}
    \\ \prod_{\ell = 1}^n \xi_T(\alpha_{\ell\mc{A}})  \eta_T(\theta_\ell)  \mel*{\alpha_{\ell\mc{A}}}{\theta_{\ell\mc{A}_1} \otimes \theta_{(\ell  + 1)\mc{A}_2} }{\alpha_{(\ell+1)\mc{A}}} 
		\, . 
    \label{eqn:trace_full_strip_geometry}
\end{multline}

To evaluate Eq.~\eqref{eqn:trace_full_strip_geometry} to an explicit function of $n$, we proceed by introducing classical variables $\sigma_s^{(\ell)}$ that take value $1$ ($0$) if the star $A_s$ is present (not present) in the decomposition of the boundary group element $\theta_\ell$. 
We can then write $\theta_\ell = \prod_s A_s^{\sigma_s^{(\ell)}}$ and
\begin{align}
    \mel*{0}{\prod_{\ell=1}^n \theta_\ell }{0}
    &= \mel*{0}{\prod_{s\in \partial} A_s^{\sum_\ell \sigma_s^{(\ell)}}}{0}
		\nonumber \\ 
		&= \prod_{s\in \partial} \delta\left( \sum_{\ell=1}^n \sigma_s^{(\ell)} \text{ mod } 2 \right) 
		\, .
    \label{eqn:sigma_variable_constraint}
\end{align}
The second equality holds because the matrix element equals unity iff an even number of $A_s$ act on each star. Although this form of the constraint is entirely valid, it is more convenient to represent the Kronecker delta as
\begin{equation}
    \delta\left( y \text{ mod } 2 \right) = \frac{1}{2} \sum_{k=0}^1 \mathrm{e}^{\mathrm{i}\pi ky} 
		\, .
\end{equation}
When we perform the product over boundary stars, the benefit of the alternative representation becomes evident: 
\begin{equation}
    \mel*{0}{\prod_\ell \theta_\ell}{0} = \frac{1}{2^{N_\partial}} \sum_{\{ k_s \}} \mathrm{e}^{\mathrm{i}\pi \sum_\ell \sum_s k_s \sigma_s^{(\ell)}} 
		\, .
    \label{eqn:sigma_variable_constraint_final}
\end{equation}
At the expense of introducing a further $N_\partial$ variables $\{ k_s \}$, we have been able to write the constraint~\eqref{eqn:sigma_variable_constraint} in a form that is separable into a product over the different replicas. The number of stars present in the decomposition of $\theta_\ell$ has a particularly simple representation in terms of $\sigma_s^{(\ell)}$ variables, $\mathscr{n}(\theta_\ell) = \sum_s \sigma_s^{(\ell)}$, which allows us to write $\eta_T(\theta_\ell)$ explicitly as
\begin{equation}
    \eta_T(\theta_\ell) = \frac{1}{2\cosh(\frac{NK_A}{2})} \sum_{J_\ell = \pm 1} \mathrm{e}^{J_\ell K_A (N/2 - \sum_s \sigma_s^{(\ell)})} 
		\, ,
\end{equation}
where we recall that $K_A = -\ln \tanh \beta\lambda_A$. 

To evaluate the $\xi_T$ factor, it is more convenient to work with configurations of the plaquette magnetisations $M_p = \ev*{B_p}_\alpha \in \{ -1, 1 \}$ rather than the physical spin configurations $\alpha$. This is possible because $\xi_T(\alpha_\mc{A}) = \xi_T(\alpha_\mc{A}^\prime)$ if all boundary plaquette magnetisations $\{ M_{p \in \partial} \}$ of $\alpha_\mc{A}$ and $\alpha_\mc{A}^\prime$ are the same, which may be shown as follows. If $M_{p}(\alpha_\mc{A}) = M_{p}(\alpha_\mc{A}^\prime)$, $\forall \, p\in \partial$, then we are able to write
\begin{equation}
    \alpha_\mc{A}^\prime = (\Gamma_\mc{A} \cdot g_\mc{A}) \, \alpha_\mc{A}
        \, ,
\end{equation}
with $\Gamma = \Gamma_\mc{A} \otimes \Gamma_\mc{B} \in \{ \mathds{I}, \Gamma_v^x \}$ and $g = g_\mc{A} \otimes g_\mc{B} \in \mc{G}$. That is, the two spin configurations $\alpha_\mc{A}$, $\alpha_\mc{A}^\prime$ differ at most by the action of a group element $g$ and the vertical winding operator $\Gamma_v^x$ (see Sec.~\ref{sec:introduction_to_model} for the definition of winding operators). This implies
\begin{equation}
    \xi_T(\alpha_\mc{A}^\prime) 
		= 
		\frac{1}{Z_B} \sum_{\alpha_B^\prime} 
		\mathrm{e}^{\beta\lambda_B M_{\bar{\partial}} \left[ \left(\Gamma_\mc{A} \cdot g_\mc{A}\right)\,\alpha_\mc{A} \otimes \alpha_\mc{B}^\prime \right]} 
		\, .
\end{equation}
We can then introduce new summation variables $\alpha_\mc{B} = \Gamma_\mc{B} g_\mc{B} \alpha_\mc{B}^\prime$, giving 
\begin{equation}
    \xi_T(\alpha_\mc{A}^\prime) = \frac{1}{Z_B} \sum_{\alpha_B} \mathrm{e}^{\beta\lambda_B M_{\bar{\partial}} \left[ \left(\Gamma \cdot g\right)\, \alpha \right]} 
		\, . 
\end{equation}
Since $M_{\bar{\partial}}\left[(\Gamma \cdot g) \, \alpha\right] = M_{\bar{\partial}}(\alpha)$, we have shown the desired result.

We can treat the magnetisations $M_p$ as Ising spins that live at the centers of the plaquettes $p$. They are subject to the condition that an even number of Ising spins are negative in any given configuration, since they must satisfy $\prod_p B_p = \mathds{1}$. If an even (odd) number of boundary plaquette magnetisations $M_{p \in \partial}$ are negative, there must be a compensating even (odd) number of negative plaquette magnetisations $M_{p \notin \partial}$ in the bulk, introducing an even--odd effect.

To evaluate $\xi_T$ we need to sum over bulk magnetisation configurations $\{ M_{p \notin \partial} \}$, keeping the boundary configuration $\{ M_{p \in \partial} \}$ fixed. This results in
\begin{equation}
    \xi_T(M_\partial) = \frac{1}{2^{3N_\partial}} \frac{\cosh(\beta\lambda_B)^{N-N_\partial} \pm \sinh(\beta\lambda_B)^{N-N_\partial}}{\cosh(\beta\lambda_B)^N + \sinh(\beta\lambda_B)^N} 
		\, . 
	\label{eqn:xi_T_loop}
\end{equation}
The upper (lower) sign corresponds to an even (odd) number of negative $M_{p\in \partial}$. The factor $1/2^{3N_\partial}$ comes from two contributions. Firstly, one must account for the fact that there are $2^{\mc{N}_\mc{A}-N_\partial}$ physical spin configurations per magnetisation configuration of the $\mc{A}$ subsystem, where $\mc{N}_\mc{A} = 3N_\partial$ is the total number of physical spins contained within $\mc{A}$, which leads to a prefactor of $1/2^{2N_\partial}$. Secondly, a factor of $1/2^{N_\partial}$ comes from the conversion of exponentials to $\sinh$ and $\cosh$.

Since each boundary star is adjacent to two boundary plaquettes, the parity of the number of negative $M_{p \in \partial}$ is conserved between replicas so that $\prod_\ell \xi_T(M_\partial(\alpha_\ell)) = \xi_T^n(M_\partial(\alpha_1))$.

Finally, we need to calculate the boundary plaquette magnetisation $M_\partial = \sum_{p \in \partial} M_p$ for each of the replicas. Given $\alpha_{1\mc{A}}$, say, then all $\alpha_{\ell\mc{A}}$ for $\ell > 1$ are uniquely determined by~\eqref{eqn:trace_full_strip_geometry}. Introducing Ising spins $\tau_s^{(\ell)} = 2\sigma_s^{(\ell)} - 1 \in \{ -1, 1 \}$, the recursion relation between boundary magnetisations $M_s$, $s = p \in \partial$,  of adjacent replicas is $M_s(\alpha_\ell) = M_s(\alpha_{\ell - 1})\tau_s^{(\ell)} \tau_{s+1}^{(\ell)} \tau_s^{(\ell -1)} \tau_{s+1}^{(\ell - 1)}$. Alternatively, in terms of $M_s(\alpha_1)$, 
\begin{equation}
    M_s(\alpha_\ell) = M_s(\alpha_1) \tau_s^{(1)} \tau_{s+1}^{(1)} \tau_s^{(\ell)}\tau_{s+1}^{(\ell)} 
		\, .
    \label{eqn:replica_magnetisation}
\end{equation}
The notation $M_s$ is interpreted as the magnetisation of the plaquette with the same index as the star $s$, as depicted in Fig.~\ref{fig:boundary_length_lattice:extended}. The description in terms of classical spin variables is now complete and we are in a position to substitute back into~\eqref{eqn:trace_full_strip_geometry}. Relabelling $\alpha_1 \to \alpha$, we obtain
\begin{widetext}
\begin{equation}
    \Tr\big(\rho_\mc{A}^{T_2}\big)^n = \frac{1}{2^{N_\partial}} \frac{1}{\left[ 2\cosh(\frac{NK_A}{2}) \right]^n} \sum_{\alpha_\mc{A}} \xi_T^n(M_\partial(\alpha_\mc{A})) \sum_{\{ k_s \}_{s=1}^{N_\partial}} \sum_{\{ J_\ell \}_{\ell = 1}^n} \prod_{\ell=1}^n \sum_{\{ \sigma_s^{(\ell)} \}_{s=1}^{N_\partial}}
    \mathrm{e}^{ -\sum_{s=1}^{N_\partial}\left[ \gamma_s \tau_s^{(\ell)}\tau_{s+1}^{(\ell)}  + h_s^{(\ell)} \sigma_s^{(\ell)}\right] + \frac12 N J_\ell K_A }
		\, ,
    \label{eqn:trace_full_strip_geometry_ising_spins}
\end{equation}
\end{widetext}
where $h_s^{(\ell)} \equiv J_\ell K_A + i\pi k_s$ and $\gamma_s \equiv -\beta\lambda_B M_s$.

\begin{figure}[b]
    \centering
    \subfloat[]{%
        \includegraphics[width=0.49\linewidth]{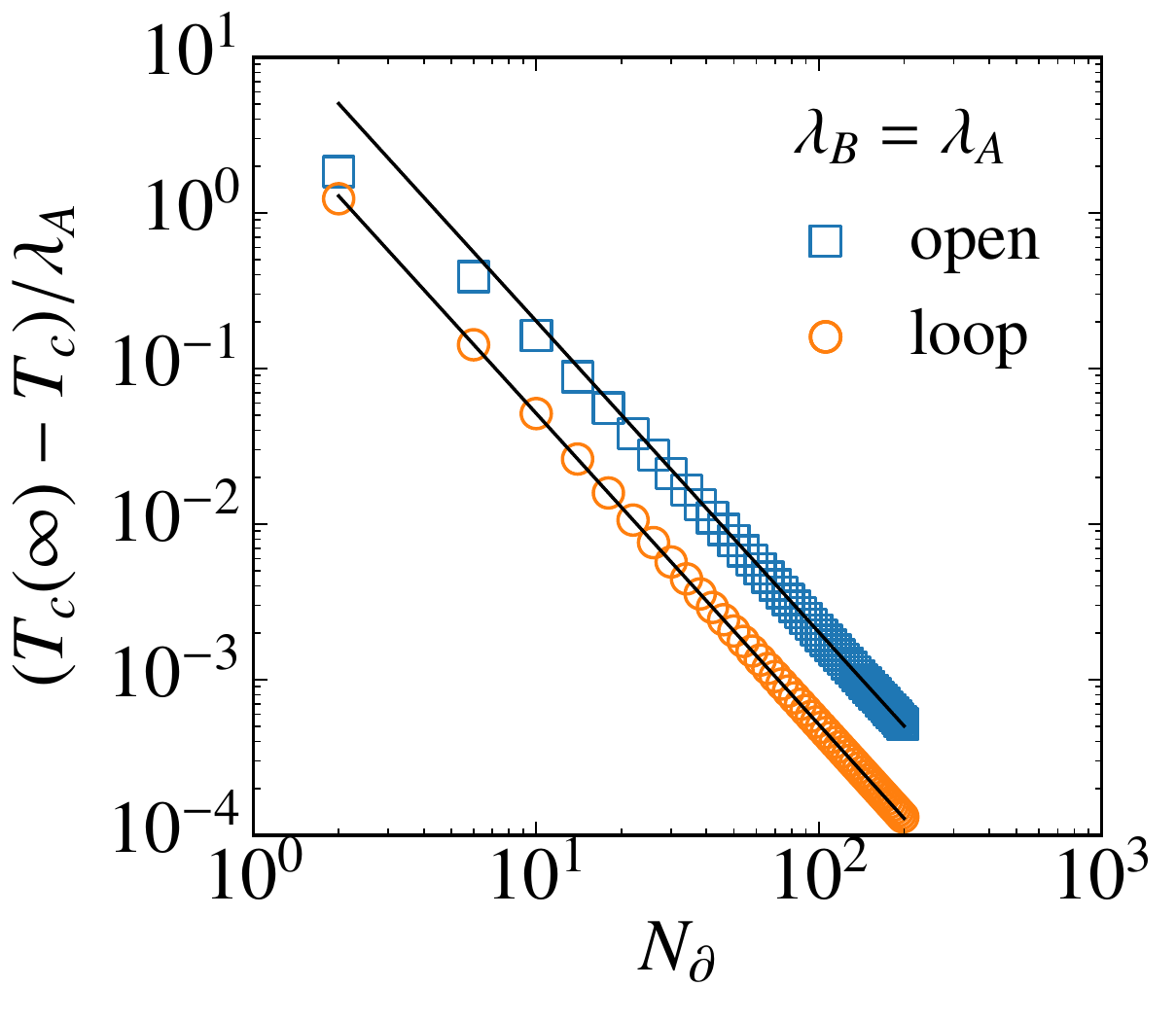}
    }%
    \hfill
    \subfloat[]{%
        \includegraphics[width=0.49\linewidth]{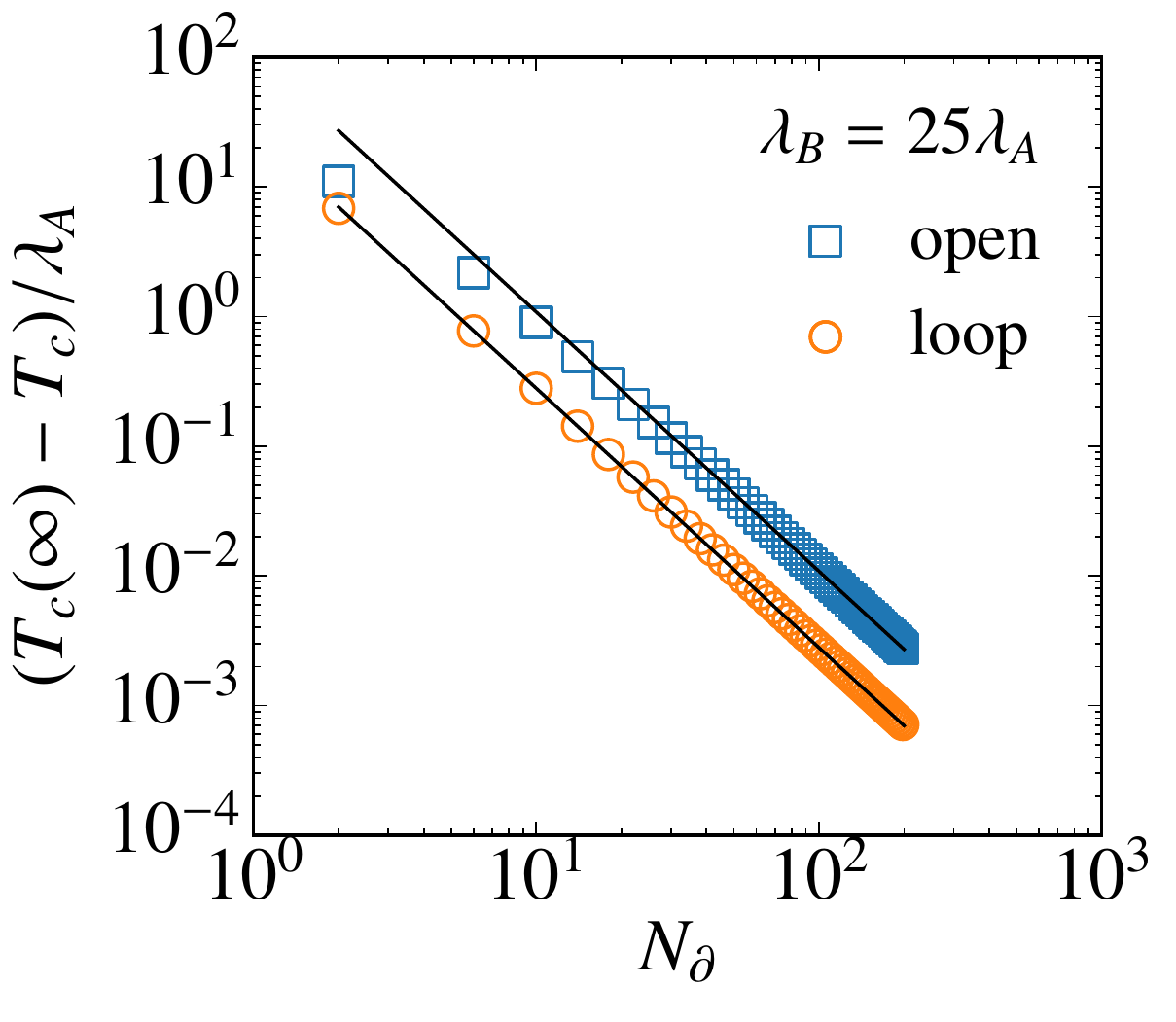}
    }%
    \caption{Convergence of the sudden death temperature $T_c$ to its limiting value $T_c(\infty) \equiv T_\infty$ for large boundaries follows a power law $T_\infty - T_c \sim N_\partial^{-2}$. Both contractible (``open'') and system-spanning (``loop'') subsystems tend towards the same limiting value.}
    \label{fig:sudden_death_limit}
\end{figure}

Notice that we have performed a relabelling of the physical spin configurations $\alpha_\mc{A}$, for each configuration $\{ \tau_s^{(1)} \}$, such that $M_s\tau_s^{(1)}\tau_{s+1}^{(1)} \to M_s$ [the parity is conserved by this transformation since $\prod_s M_s \tau_s^{(1)} \tau_{s+1}^{(1)} = \prod_s M_s $, thus leaving $\xi_T^n(M_\partial)$ unchanged]. This puts all replicas on an equal footing, making the trace separable into a product over replicas. The problem has therefore been reduced to evaluating the classical partition function
\begin{equation}
    Z(\{ \gamma_s \}, \{ k_s \}, J_\ell K_A) = \sum_{\{ \tau_s = \pm 1 \}} \mathrm{e}^{-\sum_{s=1}^{N_\partial} \left[ \gamma_s \tau_s \tau_{s+1} + \frac12 h_s^{(\ell)} (\tau_s + 1) \right] }
    \label{eqn:1D_ising_partition_function_disordered}
\end{equation}
of a 1D Ising model with reduced couplings $\{ \gamma_s \}$ in a complex reduced magnetic field $\{ h_s \}$. Using the expression
\begin{equation}
    \lim_{n\to 1} \sum_{m=0}^n \binom{n}{m} x^m y^{n-m} = \abs*{x + y} 
		\, ,
\end{equation}
having followed the even $n$ series, one can then evaluate the sum over configurations $\{ J_\ell \}$
\begin{multline}
    \sum_{\{ J_\ell \}_{\ell = 1}^n } \prod_{\ell=1}^n \mathrm{e}^{\frac12 N J_\ell K_A } Z(J_\ell K_A) 
		= \\
		\abs{\mathrm{e}^{\frac12 N K_A} Z(K_A) + \mathrm{e}^{-\frac12 N K_A} Z(-K_A)} 
		\, ,
\end{multline}
where we have suppressed the dependence of $Z$ on $\{ \gamma_s \}$ and $\{ k_s \}$ to simplify the notation. The final result follows by substituting back into~\eqref{eqn:trace_full_strip_geometry_ising_spins}:
\begin{multline}
    \norm\big{\rho_\mc{A}^{T_2}}_1 = \frac{1}{2^{N_\partial}} \frac{1}{2\cosh(\frac{NK_A}{2})} \sum_{\alpha_\mc{A}} \xi_T(M_\partial(\alpha_\mc{A})) \\ \sum_{ \{ k_s \}_{s=1}^{N_\partial} } \abs{ \mathrm{e}^{\frac12 N K_A } Z(K_A) + \mathrm{e}^{- \frac12 N K_A } Z(-K_A)  } 
		\, .
    \label{eqn:trace_norm_larger_boundary_finitesize}
\end{multline}

Taking the thermodynamic limit $N \to \infty$, the partition function $Z(-K_A)$ is suppressed by a factor $\mathrm{e}^{-NK_A}$, and the even--odd effect is removed from $\xi_T \to \sech(\beta\lambda_B)^{N_\partial}/2^{3N_\partial}$. Absorbing $\xi_T$ into the partition function as an energy shift and trading the sum over spin configurations $\alpha_\mc{A}$ for one over magnetisations $\{ M_s \}$, we obtain
\begin{equation}
    \norm\big{\rho_\mc{A}^{T_2}}_1 = \frac{1}{2^{2N_\partial}} \sum_{\{ M_s \}_{s=1}^{N_\partial}} \sum_{\{ k_s \}_{s=1}^{N_\partial}} \abs*{Z^\prime(\{ \gamma_s \}, \{ k_s \}, T)} 
		\, ,
    \label{eqn:trace_norm_larger_boundary_final}
\end{equation}
where we introduced the modified partition function
\begin{equation}
    Z^\prime = \sum_{\{ \tau_s = \pm 1 \}} \mathrm{e}^{-\sum_{s=1}^{N_\partial} \left[ \gamma_s \tau_s \tau_{s+1} + \frac12 h_s (\tau_s + 1) + \ln \cosh \beta\lambda_B \right]}
		\, .
\end{equation}
The sums in~\eqref{eqn:trace_norm_larger_boundary_final} can be regarded as performing a disorder average over the couplings and the local magnetic fields, leading to the expression $\mc{E} = \ln\ev*{\abs*{Z^\prime}}$ given in Sec.~\ref{sec:larger_boundary}.

In the limit $T \to 0^+$, one can verify after a few lines of algebra that half of the magnetisation configurations (those satisfying $\prod_s M_s = 1$), and half of the magnetic field configurations (those satisfying $\sum_s k_s \text{ mod } 2 = 0$), give a nonvanishing contribution of $2^{N_\partial + 1}$. All other configurations $\{ M_s \}$ and $\{ k_s \}$ give instead a vanishing contribution. Hence, one arrives at the stated zero-temperature result $\mc{E}(0) = (N_\partial - 1) \ln 2$, in agreement with Refs.~\onlinecite{Castelnovo2013, Lee2013}.

If the boundary does not span the entire system, the calculation must be adjusted to account for the open ends of the boundary. This results in a slightly more involved calculation, the details of which we hint at in Appendix~\ref{sec:extended_contractible_boundary}. One arrives at an expression which displays qualitatively similar behaviour to the system-spanning strip. The agreement becomes quantitative in the limit of large boundaries, where both system-spanning and contractible boundaries give the same sudden death temperature $T_\infty$, as evidenced by Fig.~\ref{fig:sudden_death_limit}. As one would expect, the limiting case $N_\partial = 1$ reproduces exactly the SPP expression of Sec.~\ref{sec:single_star_calculation}.
%
%


\subsection{$\lambda_B \to \infty$ limit}
\label{sec:infinite_lambda_B_calculation}

We now consider a limiting case that allows for a substantial simplification of the calculation of the negativity, and tells us something interesting about its behaviour at finite temperature in the presence of hard constraints. 
Namely, we take the limit $\lambda_B \to \infty$ that removes magnetic defects from the system at all temperatures. 
We also make the following working assumptions about the partition of the system: (i) $\mc{A}$ is contractible, i.e.,~not system-spanning in either direction, and (ii) for each boundary star there corresponds at least one boundary plaquette.

If these assumptions are satisfied, the matrix elements in~\eqref{eqn:trace_full_after_redefinition} imply that $\theta_\ell = \theta_{\ell+1} \equiv \theta$, $\forall \, \ell$. This is because only terms for which $M(\alpha_\ell) = N$, $\forall \, \ell$, contribute (due to $Z_B^n$ diverging as $\mathrm{e}^{nN\beta\lambda_B}$ for $\beta\lambda_B \gg 1$) and we then must have $\theta_{\ell\mc{A}_1} \otimes \theta_{(\ell + 1)\mc{A}_2} \in \mc{G}_{\mc{A}_1\mc{A}_2}$ to satisfy $M(\alpha_{\ell+1})=M(\alpha_\ell)$. 

Given $\alpha_1 \equiv \alpha$, say, then each $\alpha_{\ell\mc{A}}$ for $\ell > 1$ is uniquely determined by the matrix elements in~\eqref{eqn:trace_full_after_redefinition}: $\ket*{\alpha_{(2m+1)\mc{A}}} = \ket*{\alpha_\mc{A}}$ and $\ket*{\alpha_{2m\mc{A}}} = \theta \ket*{\alpha_\mc{A}}$ for integer $m$. Since $n$ is assumed even, the constraint $\prod_\ell \theta_\ell = \theta^n = \mathds{I}$ is automatically satisfied. On the other hand, the $\alpha_{\ell\mc{B}}$ for $\ell > 1$ are only subject to the condition that their plaquette magnetisation remains undegraded from $M(\alpha_\ell) = N$, and therefore each $\alpha_\ell$ summation, $\ell > 1$, produces a factor of $4\abs*{\mc{G}_\mc{B}}$.
The remaining summation over $\alpha$, in the limit $\lambda_B \to \infty$, gives instead 
\begin{equation}
    \frac{1}{Z_B^n} \sum_{\alpha} \mathrm{e}^{n\beta\lambda_BM(\alpha)} 
		\sim 
		\frac{\sum^\prime_\alpha \mathrm{e}^{n N \beta\lambda_B} }
		     {\left(\sum^\prime_\alpha \mathrm{e}^{N \beta\lambda_B} \right)^n} \to \frac{1}{[4\abs*{\mc{G}}]^{n-1}} 
		\, ,
\end{equation}
where the prime on the summation denotes that the constraint $M(\alpha) = N$ is imposed on the spin configurations $\alpha$. 
We therefore arrive at the expression 
\begin{multline}
    \Tr\big(\rho_\mc{A}^{T_2}\big)^n = 
    \frac{\abs*{\mc{G}_\mc{B}}^{n-1}}{\abs*{\mc{G}}^{n-1}}
		\sum_{\substack{\bar{g}_1, \ldots, \bar{g}_n \\ \in \mc{G}_{\mc{A}_1}}} \sum_{\substack{\bar{\bar{g}}_1, \ldots, \bar{\bar{g}}_n \\ \in \mc{G}_{\mc{A}_2}}} \sum_{\theta \in \mc{G}_{\mc{A}_1\mc{A}_2}} 
		\times \\ 
    \mel*{0}{\prod_{\ell=1}^n \bar{g}_\ell \bar{\bar{g}}_\ell }{0}
    \left( \prod_{\ell = 1}^n \eta_T(\bar{g}_\ell\bar{\bar{g}}_\ell \theta) \right) 
		\, .
\end{multline}

Before proceeding further, we evaluate the zero-temperature limit and compare with earlier results in Ref.~\onlinecite{Lee2013}. When $T \to 0^+$, $\eta_T(g) \to 1$ $\forall \, g \in \mc{G}$, and one simply needs to count the number of terms in the summation, leading to
\begin{equation}
    \Tr\big(\rho_\mc{A}^{T_2}\big)^n
		= 
		\abs*{\mc{G}_{\mc{A}_1\mc{A}_2}}\left[ \frac{\abs*{\mc{G}_{\mc{A}_1}}\abs*{\mc{G}_{\mc{A}_2}}\abs*{\mc{G}_\mc{B}}}{\abs*{\mc{G}}} \right]^{n-1}.
\end{equation}
One must remember that we arrived at this expression following the even $n$ series. Taking $n \to 1$ and decomposing the number of stars as $N_\mc{A} = N_{\mc{A}_1} + N_{\mc{A}_1} + N_\partial$, we arrive at $\mc{E}(0) = N_\partial \ln 2$, in agreement with Ref.~\onlinecite{Lee2013}. Here $N_{\mc{A}_i}$ corresponds to the number of stars that act solely on subsystem $\mc{A}_i$ while $N_\partial$ equals the number of stars that straddle the $\mc{A}_1$--$\mc{A}_2$ boundary. Had we followed the odd series, the constraint $\prod_\ell \theta_\ell = \theta = \mathds{I}$ removes the summation over $\theta$ and we would obtain instead $\lim_{n \to 1} \Tr\big(\rho_\mc{A}^{T_2}\big)^n = 1$, as required for the trace of a density matrix. 

Taking $n\to 1$ for finite $T$ is more difficult. In order to carry out the calculation, in a similar manner to Sec.~\ref{sec:larger_boundary_calculation}, we need to introduce $2n+1$ sets of Ising spins $\{\vartheta_s^{(\ell)}\}$, $\{\varphi_s^{(\ell)}\}$ and $\{\psi_s\}$. The spins are defined as follows: $\vartheta_s^{(\ell)} = -1(1)$ if $A_s$ appears (does not appear) in the decomposition of $\bar{g}_\ell \in \mc{G}_{\mc{A}_1}$, and similarly for $\varphi_s^{(\ell)}$ and $\psi_s$ with respect to $\bar{\bar{g}}_\ell \in \mc{G}_{\mc{A}_2}$ and $\theta \in \mc{G}_{\mc{A}_1\mc{A}_2}$. 
For convenience, we define
\begin{equation}
    \Delta_\psi \equiv \frac{N - N_\mc{A}}{N_\partial} > 1
        \, ,
\end{equation}
and introduce the notation $\tilde{\psi}_s = \psi_s + \Delta_\psi$, so that in terms of the spin variables, the number of stars flipped by the action of the group element $\bar{g}_\ell \bar{\bar{g}}_\ell \theta$ can be written as 
\begin{equation}
    \mathscr{n}(\bar{g}_\ell \bar{\bar{g}}_\ell \theta) = \frac{N}{2} - \frac{1}{2}\left( \sum_{s \in \mc{A}_1}\vartheta_s^{(\ell)} + \sum_{s \in \mc{A}_2}\varphi_s^{(\ell)} + \sum_{s \in \partial}\tilde{\psi}_s \right) 
		\, .
    \label{eqn:ising_spins_n}
\end{equation}
The constraint imposed by the matrix element $\bra{0} \prod_\ell \bar{g}_\ell \bar{\bar{g}}_\ell \ket{0} \neq 0$, which is satisfied iff an even number of star operators acts on each site, becomes 
\begin{equation}
    \prod_{s \in \mc{A}_1} \delta\left( \prod_{\ell = 1}^n \vartheta_s^{(\ell)} - 1 \right) \prod_{s \in \mc{A}_2} \delta\left( \prod_{\ell = 1}^n \varphi_s^{(\ell)} - 1 \right)
		\, .
    \label{eqn:ising_constraint}
\end{equation}

Denoting a configuration of spins, for a given $\ell$, by $ \chi_\ell = \{ \vartheta_s^{(\ell)}, \varphi_s^{(\ell)}, \psi_s \}$ we can complete the transformation to the Ising spin language by writing
\begin{multline}
    2 \cosh\left( \frac{K_A}{2} N \right) \sum_{\{ \vartheta_s^{(\ell)}, \varphi_s^{(\ell)} \}} \eta_T (\chi_\ell) =  \\ \sum_{J_\ell = \pm 1} \left( \prod_{s \in \mc{A}_1} \sum_{\vartheta_s^{(\ell)} = \pm 1} \mathrm{e}^{(K_A/2)J_\ell \vartheta_s^{(\ell)}} \right) \left( \prod_{s \in \mc{A}_2} \sum_{\varphi_s^{(\ell)} = \pm 1} \mathrm{e}^{(K_A/2)J_\ell \varphi_s^{(\ell)}} \right) \\ \left( \prod_{s \in \partial} \mathrm{e}^{(K_A/2)J_\ell \tilde{\psi}_s} \right).
    \label{eqn:ising_spins_eta_summed}
\end{multline}
Collecting together results from~\eqref{eqn:ising_spins_n} to~\eqref{eqn:ising_spins_eta_summed}, one obtains an explicit expression for the trace in terms of the Ising spin variables
\begin{widetext}
    \begin{multline}
            \Tr\big(\rho_\mc{A}^{T_2}\big)^n 
    				= 
    				\frac{\abs*{\mc{G}_\mc{B}}^{n-1}}{\abs*{\mc{G}}^{n-1}} \frac{1}{\left[2 \cosh\left( \frac{K_A}{2} N \right) \right]^n} 
    				\: \times \\
    				\sum_{\{ J_\ell \}_{\ell=1}^n} \sum_{\{\psi_s \}_{s=1}^{N_\partial}} 
             \left( \prod_{s \in \mc{A}_1} \sum_{\substack{\{ \vartheta_s^{(\ell)} \}_{\ell=1}^n\\ \prod_\ell \vartheta_s^{(\ell)} = +1}} \mathrm{e}^{(K_A/2)\sum_\ell J_\ell \vartheta_s^{(\ell)}} \right) \left( \prod_{s \in \mc{A}_2} \sum_{\substack{\{ \varphi_s^{(\ell)} \}_{\ell=1}^n\\ \prod_\ell \varphi_s^{(\ell)} = +1}} \mathrm{e}^{(K_A/2)\sum_\ell J_\ell \varphi_s^{(\ell)}} \right) \left( \prod_{s \in \partial} \mathrm{e}^{(K_A/2)\sum_\ell J_\ell\tilde{\psi}_s} \right)
    				 \, .
             \label{eqn:trace_full_infinite_lambda_B}
    \end{multline}
\end{widetext}
The explicit constraint has now been replaced by conditions on the relevant summations and we have exchanged the order of the product over $\ell$ with the summations over $J_\ell$ using $\prod_{\ell} (\sum_{J_\ell = \pm 1} F(J_\ell)) = \sum_{\{ J_\ell \}} \prod_{\ell} F(J_\ell)$, for arbitrary $F$.

The sums over $\{ \vartheta_s^{(\ell)} \}$ and $\{ \varphi_s^{(\ell)} \}$ can be carried out explicitly by exploiting their analogy to the partition function of a 1D Ising model as follows. First, note that the site labels $s$ of $\vartheta_s^{(\ell)}$ and $\varphi_s^{(\ell)}$ are mute since the contribution from every star $s \in \mc{A}_1$ is identical, and similarly for every star $s \in \mc{A}_2$. This allows us to substantially simplify the notation $\vartheta_s^{(\ell)} \to \vartheta^{(\ell)}$ and $\varphi_s^{(\ell)} \to \varphi^{(\ell)}$. Then we make use of the one-to-two mapping $\vartheta^{(\ell)} = \tau_\ell \tau_{\ell+1}$, and similarly for the $\varphi^{(\ell)}$ spins. The motivation behind this mapping is that the constraint $\prod_\ell \vartheta^{(\ell)} = +1$ is automatically satisfied if we impose periodic boundary conditions, $\tau_{n+1} = \tau_1$, since $\tau_\ell^2 = 1$. A factor of $1/2$ is required to compensate for the double counting of spin configurations in the $\tau$ spin notation,
\begin{align}
    \sum_{\substack{\{ \vartheta^{(\ell)} \}_{\ell=1}^n\\ \prod_\ell \vartheta^{(\ell)} = +1}} \mathrm{e}^{(K_A/2)\sum_\ell J_\ell \vartheta^{(\ell)}} &= \frac{1}{2} \sum_{\{ \tau_\ell \}_{\ell=1}^n} \mathrm{e}^{(K_A/2)\sum_\ell J_\ell \tau_\ell \tau_{\ell + 1}} 
		\nonumber \\ 
		&\equiv \frac12 Z_n^p(K_A, \{ J_\ell \})
		\, .
\end{align}
This is nothing but the partition function of a 1D Ising chain of length $n$, with position-dependent reduced couplings $K_A J_\ell / 2$, which we denote for convenience as $Z_n^p$ (the $p$ reminding us of the periodic boundary conditions). A straightforward manipulation (see for instance Ref.~\onlinecite{Castelnovo2007}) gives us 
\begin{equation}
    Z_n^p(K_A, \{ J_\ell \}) = \left[ 2\cosh\left( \frac{K_A}{2} \right) \right]^n + \left( \prod_{\ell = 1}^n J_\ell \right) \left[ 2\sinh\left( \frac{K_A}{2} \right) \right]^n 
		\, .
\end{equation}
Substituting back into~\eqref{eqn:trace_full_infinite_lambda_B}, we are hence left with the simpler expression
\begin{multline}
    \Tr\big(\rho_\mc{A}^{T_2}\big)^n 
		= 
		\frac{|\mc{G}_\mc{B}|^{n-1}}{|\mc{G}|^{n-1}} \dfrac{1}{\left[ 2\cosh\left( \frac{K_A}{2} N \right) \right]^n} \sum_{\{\psi_s \}_{s=1}^{N_\partial}} \\ \sum_{\{ J_\ell \}_{\ell=1}^n} \left[ \tfrac12  Z_n^p\right]^{N_{\mc{A}_1} + N_{\mc{A}_2}}  \mathrm{e}^{(K_A / 2) \sum_\ell J_\ell \sum_s \tilde{\psi}_s} 
		\, .
    \label{eqn:trace_partition_functions}
\end{multline}
To obtain an explicit function of $n$, as is required to take the limit $n \to 1$, we must perform the summation over configurations $\{ J_\ell \}$.  Thankfully, the summand in~\eqref{eqn:trace_partition_functions} only depends on $J_\ell$ though $\prod_\ell J_\ell$ and $\sum_\ell J_\ell$. We can take advantage of this by trading the sum over configurations $\{ J_\ell \}_{\ell = 1}^n$ for one over the number of negative $J_\ell$, denoted by $m$, with the appropriate combinatoric weighting $\binom{n}{m}$. The terms on the second line in~\eqref{eqn:trace_partition_functions} become
\begin{multline}
    \sum_{\substack{m=0 \\ \mathrm{even}}}^n \binom{n}{m} \left[ \tfrac12 Z_n^+ \right]^{N_\mc{A} + N_\mc{B}} \mathrm{e}^{(K_A/2)\sum_s \tilde{\psi}_s(n- 2m)} 
		\: + \\
    \sum_{\substack{m=0 \\ \mathrm{odd}}}^n \binom{n}{m} \left[ \tfrac12 Z_n^- \right]^{N_\mc{A} + N_\mc{B}} \mathrm{e}^{(K_A/2)\sum_s \tilde{\psi}_s(n- 2m)}
		\, .
    \label{eqn:negativity_sum_over_m}
\end{multline}
With the help of the binomial generating function $\sum_{m=0}^n \binom{n}{m} x^m = (1 + x)^n$, evaluated at $x$ and $-x$, we can then evaluate the sums over even and odd $m$,
\begin{align}
    \lim_{n \to 1} \sum_{\substack{m=0 \\ \mathrm{even}}}^n \binom{n}{m} x^m &= \frac12 \Big( \abs{1 + x} + \abs{1-x} \Big) = \max(1,x), \label{eqn:binomial_identity1}\\
    \lim_{n \to 1} \sum_{\substack{m=0 \\ \mathrm{odd}}}^n \binom{n}{m} x^m &= \frac12 \Big( \abs{1 + x} - \abs{1-x} \Big) = \min(1, x)
		\, ,
		\label{eqn:binomial_identity2}
\end{align}
where we used the fact that $(1 \pm x)^n \to \abs{1 \pm x}$, for $n \to 1$ following the even $n$ series. Note that the Ising partition function has the limit $\tfrac12 Z_{n \to 1}^\pm \to \mathrm{e}^{\pm  (K_A / 2)}$. It is crucial at this stage to remember that $\tilde{\psi}_s > 0$, so that~\eqref{eqn:negativity_sum_over_m} evaluates to 
\begin{equation}
    \sum_{\sigma = \pm 1} \mathrm{e}^{\sigma(K_A / 2) (N_{\mc{A}_1} + N_{\mc{A}_2} + \sum_s \tilde{\psi}_s)}
		\, .
\end{equation}
Both the remaining sums over $\{ \psi_s \}$ and over $\sigma$ reduce to the partition function of noninteracting Ising spins in a magnetic field, leading to the main result of this section
\begin{equation}
    \mathrm{e}^{\mc{E}(T)} = \frac{\cosh\left(\frac12 (N - N_\partial) K_A \right)}{\cosh\left(\frac12 N K_A\right)} \left[ 2\cosh\left(\frac{K_A}{2}\right) \right]^{N_\partial}
        \, ,
\end{equation}
as discussed in Sec.~\ref{sec:infinite_lambda_B}.


\section{Experimental Considerations}
\label{sec:experimental_considerations}

Although the negativity remains an invaluable theoretical tool, allowing us to efficiently understand and quantify the entanglement content of mixed quantum states, it is not a quantity that is directly accessible in experiments.
However, it is possible to obtain the negativity experimentally via explicit reconstruction of the system's density matrix using quantum state tomography~\cite{Vogel1989}. Such an approach is hindered by the exponential scaling in the required resources with system size, which prevents tomography from being applied to larger systems. This shortcoming can be addressed in weakly entangled systems, where more efficient tomography schemes with polynomial scaling exist~\cite{Cramer2010, Lanyon2017}.
Alternatively, it is possible to approximate the negativity by measuring moments of the partially transposed density matrix in a scheme recently proposed by Gray \emph{et al.}~\cite{Gray2017}. This method also exhibits polynomial scaling with system size.

In this manuscript, we have worked exclusively with Kitaev's toric code.
Despite being highly artificial, vast interest in the topological properties of this model, particularly with reference to quantum computation, has led to the proposition of various experimental implementations of the Hamiltonian~\cite{Weimer2010, Weimer2011, Sameti2017} and its ground states~\cite{Han2007, Aguado2008, Lu2009, Pachos2009}. The latter approach has proved more fruitful, since directly simulating four-body interactions is experimentally challenging. An alternative method, which only requires the simulation of two-body interactions, consists of cold atoms hopping on a honeycomb lattice~\cite{Duan2003}. In this setup, the toric code Hamiltonian appears as a perturbative limit~\cite{Kitaev2006}.


\section{conclusions}
\label{sec:conclusions}

We performed exact calculations of the logarithmic negativity at finite temperature in the 2D toric code. The exact formulae that we have presented in this manuscript have simple and intuitive physical interpretations that provide us with a deeper understanding of finite temperature entanglement and sudden death in two-dimensional lattice systems.

The smallest nontrivial choice of subsystem---a neighbouring star plaquette pair (SPP)---allows us to extract the general behaviour of the negativity at finite temperatures. We establish that, for this subsystem, the negativity is both a necessary and sufficient condition for separability. Below the gap to the least energetically costly elementary excitations (defects), the negativity remains exponentially close to its zero-temperature value. Above this gap, defects are thermally excited with some $\mc{O}(1)$ density, and the negativity correspondingly decays. Sudden death of the negativity, and therefore entanglement, occurs at a temperature of order the largest energy scale in the problem, consistent with previous studies in other models, e.g.,~Refs.~\onlinecite{Anders2008, Sherman2016}. This corresponds to the thermal energy required to excite an $\mc{O}(1)$ density of the most energetically costly defect.
We have verified that this physical picture works well for other systems, including the harmonic lattice and Ising chain, which are discussed in Appendices~\ref{sec:ising_model} and \ref{sec:harmonic_lattice}, respectively.

Our main results concern the extension of the SPP to a system consisting of a block of stars connected to a block of plaquettes by a boundary of length $N_\partial > 1$ stars. We find that extending the length of the boundary leads to some important differences. The number of discontinuities in the first derivative of $\mc{E}$, which arise whenever an eigenvalue of the partially transposed reduced density matrix changes sign, increases exponentially with $N_\partial$. Sudden death of the negativity is pushed to higher temperatures, tending towards approximately twice the SPP value in the limit of large boundaries. Therefore a higher density of magnetic defects is required to kill entanglement between subsystems that share a thermodynamically large boundary, but the temperature required to create this density is still $\mc{O}(\lambda_B)$ (up to logarithmic factors).
We have also shown that, for sufficiently large boundaries, the negativity no longer vanishes abruptly at $T_c$; instead it vanishes with zero gradient asymptotically in the thermodynamic limit.

The interpretation of the sudden death temperature $T_c$ for subsystems larger than the SPP is less clear-cut, as the negativity is only a necessary condition for separability. It represents a lower bound for the temperature above which \emph{all} entanglement vanishes, including any quantum correlations not picked up by the negativity. Any entanglement that may be present in the system above $T_c$ is, however, not able to be distilled, and as a result is often referred to as \emph{bound entanglement}~\cite{Ferraro2008}. We would welcome the introduction of an entanglement monotone that is both necessary \emph{and} sufficient for separability, but expect on physical grounds that our conclusions would remain unchanged up to $\mc{O}(1)$ factors.

Finally, it was shown that imposing the hard constraint $\lambda_B \to \infty$ on the system leads to nonzero entanglement at all temperatures. The constraint renders some portion of the full Hilbert space thermally inaccessible, i.e.,~the span of the states with a nonzero number of magnetic defects, thereby locking in the zero-temperature magnetic loop structure. We observe a slow $1/T$ demise of the negativity up to arbitrarily large temperatures as thermal fluctuations gradually wash out the zero-temperature quantum correlations. Only at infinite temperature, where the density matrix is proportional to the identity operator, is the system completely separable. One can carry out very similar calculations to obtain the negativity of a hard-constrained bipartite system, but the results are no more insightful and were therefore omitted.
This provides a potentially physical example of how a hard constraint applied to a system with a macroscopically degenerate projected manifold can result in thermally robust entanglement. This is likely to be the case for instance in other spin liquids and frustrated magnetic systems in general, when projected down to their low-energy states.


\section*{Acknowledgements}

We are grateful to Pasquale Calabrese for bringing this problem to our attention in the first place. We also thank him, as well as Claudio Chamon and Attila Szab\'{o}, for useful discussions. This work was supported in part by the Engineering and
Physical Sciences Research Council (EPSRC) Grants No.~EP/G049394/1 and No.~EP/M007065/1.

\appendix


\section{Defect density}
\label{sec:defect_density}

The defect density of each species $X=A, B$ can be calculated from the expression
\begin{equation}
    \varrho_X = \frac{1}{N} \pdv{\ln Z_X}{\epsilon}
        \, ,
    \label{eqn:defect_density_calculation}
\end{equation}
where we have introduced $\epsilon \equiv 2\beta\lambda_X$ and the partition function 
\begin{equation}
    Z_X = \sum_{\substack{ n=0 \\ \text{even} }}^N \binom{N}{n} \mathrm{e}^{-n\epsilon}
        \, .
\end{equation}
The degeneracy of each energy level has been omitted since it cancels in \eqref{eqn:defect_density_calculation}. Evaluating Eq.~\eqref{eqn:defect_density_calculation}, we arrive at the full expression
\begin{equation}
    \varrho_X = \mathrm{e}^{-2\beta\lambda_X} \frac{  (1 + \mathrm{e}^{-2\beta\lambda_X} )^{N-1} - (1 - \mathrm{e}^{-2\beta\lambda_X} )^{N-1}} { (1 + \mathrm{e}^{-2\beta\lambda_X})^N + (1 - \mathrm{e}^{-2\beta\lambda_X})^N }
        \, ,
\end{equation}
which, in the thermodynamic limit $N \to \infty$, reduces to
\begin{equation}
    \varrho_X \to \frac{1}{1 + \mathrm{e}^{2\beta\lambda_X}} \equiv n_\text{F}(2\lambda_X)
        \, ,
\end{equation}
i.e.,~the Fermi-Dirac distribution. The constraint that electric and magnetic defects must be created in pairs has a vanishing effect in the thermodynamic limit---as far as the density is concerned, each star (plaquette) behaves as an independent two-level system with energy gap $\lambda_A$ ($\lambda_B$).


\section{Ising model}
\label{sec:ising_model}

A brief study of the one-dimensional transverse field Ising model (TFIM) highlights some of the peculiarities associated with the toric code. The Hamiltonian of an Ising chain with $L$ sites and periodic boundary conditions is
\begin{equation}
    H_L = -J \sum_{i=1}^L \sigma_i^z \sigma_{i+1}^z - h\sum_{i=1}^L \sigma_i^x
        \, ,
    \label{eqn:ising_hamiltonian}
\end{equation}
where $J$ ($h$) controls the strength of nearest-neighbour interactions (magnetic field).
For the limiting case of two sites, the Hamiltonian~\eqref{eqn:ising_hamiltonian} can be written in matrix form
\begin{equation}
    H_2 = -
    \begin{pmatrix}
         J &  h &  h &  0 \\
         h & -J &  0 &  h \\
         h &  0 & -J &  h \\
         0 &  h &  h &  J
    \end{pmatrix}
        \, ,
\end{equation}
with respect to the eigenstates of $\otimes_i \sigma_i^z$, and has energy eigenvalues $\pm \epsilon_1 = \pm J$ and $\pm \epsilon_2 = \pm \sqrt{J^2 + 4 h^2}$. The density matrix $\rho = \mathrm{e}^{-\beta H}/Z$ and its partial transpose (over the second site) can be evaluated explicitly to give the negativity
\begin{equation}
    \mathrm{e}^{\mc{E}(T)} = \frac{2}{Z} \left[ \cosh(\beta\epsilon_2) + \max \left( \cosh(\beta\epsilon_1), \frac{\epsilon_1}{\epsilon_2} \sinh(\beta\epsilon_2) \right) \right]
        \, ,
\end{equation}
with the partition function $Z = 2\cosh(\beta\epsilon_1) + 2\cosh(\beta\epsilon_2)$.

In the limit of zero temperature, $ T \to 0^+$, we obtain
\begin{equation}
    \mc{E}(0) = \ln( 1 + \frac{J}{\sqrt{J^2 + 4 h^2}} )
        \, ,
    \label{eqn:2_site_ising_zeroT_negativity}
\end{equation}
for nonzero field $h$, whereas the negativity vanishes identically if $h=0$ from the outset.
In contrast with the toric code, the zero-temperature negativity is a function of the Hamiltonian's parameters. The expression \eqref{eqn:2_site_ising_zeroT_negativity} is maximal ($\mc{E} = \ln 2$) for infinitesimal field when the ground state coincides with one of the maximally entangled Bell states $\ket*{\Phi^+} = (\ket*{00}+\ket*{11})/\sqrt{2}$. Increasing the field strength makes it increasingly favourable for the spins to align with the field, eventually leading to $\mc{E}(0) \to 0$ in the limit of $h/J \to \infty$ where the ground state, $\ket{\sigma_i^x = +1}$, is separable.

Similarly to the toric code, the sudden death temperature is set by the larger of the two energy scales (up to logarithmic factors)
\begin{equation}
    T_c \simeq
    \begin{cases}
        \dfrac{2J}{W\left(\frac{8J^2}{\mathrm{e}^2 h^2}\right)} \: &\text{ for } J \gg h \, , \\
        \dfrac{2h}{\ln\left( \frac{4h}{J} \right) } \: &\text{ for } h \gg J \, .
    \end{cases}
\end{equation}
However, the toric code and TFIM behave very differently at temperatures $T < T_c$. If one takes $J \gg h$ in an attempt to preserve the largest zero-temperature entanglement $\mc{E}(0) \simeq \ln 2$, one arrives at $\mc{E}(T) \simeq h^2/(JT)$ for $h^2/J \ll T \ll T_c$. That is, for any given nonzero temperature, the negativity vanishes in the limit $J \to \infty$. The slow $1/T$ decay of the negativity that was observed in the toric code is suppressed by a factor of $h/J$. This is because the energy gap $\varDelta$ to the first excited state in the TFIM vanishes as $\varDelta \simeq 2h^2/J$, while it remains finite in the toric code.

This effect is made more significant when the length $L$ of the chain is increased, since the energy gap becomes exponentially small in $L$, i.e.,~$\varDelta \sim h (h/J)^{L-1}$. Taking the thermodynamic limit $L \to \infty$ and then temperature to zero $T \to 0^+$, the negativity of the infinite chain vanishes like $\mc{E}(0) \sim J / h$ for $h \gg J$ and $\mc{E}(0) \sim h^2/J^2$ for $J \gg h$. Therefore, in the infinite 1D TFIM, the divergence of the sudden death temperature is accompanied by the vanishing of the zero-temperature negativity. The toric code is special in that its ground state(s), and therefore its zero-temperature negativity, are independent of the system's parameters. It is this feature that allows quantum correlations to survive up to arbitrarily high temperatures when one takes $\lambda_B \to \infty$.


\section{Extended, contractible boundary}
\label{sec:extended_contractible_boundary}

When the boundary is not system spanning, one must change the magnetisation recursion relations from Eq.~\eqref{eqn:replica_magnetisation} to
\begin{equation}
M_s(\alpha_\ell) = M_s(\alpha_{\ell - 1}) 
    \begin{cases}
        \tau_s^{(\ell)} \tau_{s+1}^{(\ell)} \tau_s^{(\ell -1)} \tau_{s+1}^{(\ell - 1)} &\text{ for } s < N_\partial\, ,\\ \tau_s^{(\ell)} \tau_s^{(\ell - 1)} &\text{ for } s = N_\partial 
				\, .
    \end{cases}
    \label{eqn:modified_replica_magnetisation}
\end{equation}
These relations embody the fact that boundary plaquettes $p=s < N_\partial$ are adjacent to two boundary stars, while the plaquette at the end of the boundary, $p=s=N_\partial$, is adjacent to only one boundary star.
As long as we work in the thermodynamic limit $N \to \infty$, so that the even--odd effect is removed from $\xi_T$ in Eq.~\eqref{eqn:xi_T_loop}, all other aspects of the calculation remain unchanged with respect to the system-spanning case.
The effect of the relations~\eqref{eqn:modified_replica_magnetisation} is to change the ``Hamiltonian'' in $Z^\prime = \sum_{\{ \tau_s \}} \mathrm{e}^{-H}$ to
\begin{multline}
    H = \gamma_{N_\partial}\tau_{N_\partial} + \sum_{s < N_\partial} \gamma_s \tau_s \tau_{s+1} \\ +\sum_s \left[ \tfrac12 h_s (\tau_s + 1) + \ln \cosh \beta\lambda_B \right] 
		\, .
\end{multline}

The SPP result is reobtained for the special case $N_\partial = 1$ in the above expression. Since the boundary is not system spanning, the zero-temperature negativity is given by $\mc{E}(0) = N_\partial \ln 2$. The partition function with the largest zero, thereby giving rise to sudden death, corresponds to the disorder realisation $k_s = 1$ $\forall \, s$ and $\gamma_s = \beta\lambda_B$ for $s < N_\partial$ and $M_{N_\partial} = -\beta\lambda_B$. This makes the equivalent of Eq.~\eqref{eqn:sudden_death_condition_larger_boundary}, which defines $T_c$ implicitly, more complex and therefore less transparent.


\section{Harmonic lattice}
\label{sec:harmonic_lattice}

\begin{figure}
    \centering
    \includegraphics[width=0.85\linewidth]{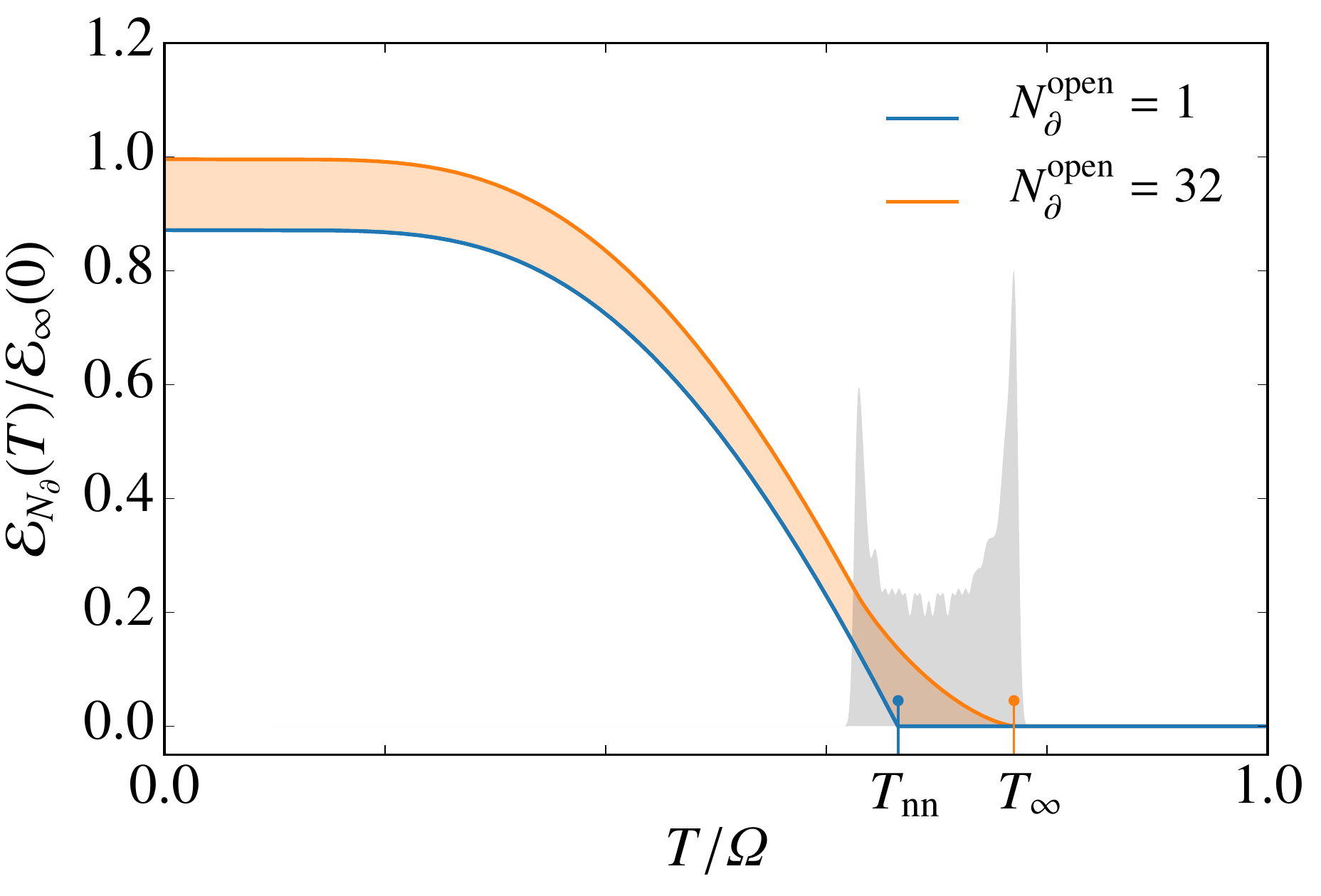}
    \caption{Negativity for a contractible strip partition (i.e.,~width two sites) in a harmonic lattice with $\varOmega = \omega$. The vertical lines correspond to $T_c$ for a nearest-neighbour pair ($T_\text{nn}$) and for an infinite boundary ($T_\infty$). The density of discontinuities in gray exhibits bunching at its edges.}
    \label{fig:harmonic_lattice_negativity}
\end{figure}

In order to test the generality of our results, we compute the negativity in the harmonic lattice numerically. We consider $N = N_\mc{A} + N_\mc{B}$ harmonic oscillators arranged on a two-dimensional square lattice with the Hamiltonian
\begin{equation}
    H = \sum_i \frac{1}{2M} p_i^2 + \frac{1}{2}M \omega^2 u_i^2 + \sum_{\langle i, j \rangle} \frac{1}{2} K (u_i - u_j)^2
        \, ,
\end{equation}
where the momenta $p_i$ and displacements $u_i$ satisfy the usual commutation relations $\comm*{u_i}{p_j} = \mathrm{i}\delta_{ij}$. The strength of the confining potential at each site (nearest-neighbour coupling) is parameterised by $\omega$ ($K$), and the mass of each oscillator is $M$. The negativity can be computed efficiently using the correlation matrix techniques outlined in Refs.~\onlinecite{Audenaert2002, Anders2008, Calabrese2015, DeNobili2016}. Given the correlation matrices $\mathds{Q}_{ij} = \ev*{u_i u_j}_\beta$ and $\mathds{P}_{ij} = \ev*{p_i p_j}_\beta$ with $i, j \in \mc{A}$, the negativity is given in terms of the positive eigenvalues $\{ \nu_1^2, \ldots, \nu_{N_{\mc{A}}}^2 \}$ of the matrix $\mathds{Q}\cdot \mathds{P}^{T_2}$. Partial transposition is effected by reversing the momenta of the oscillators belonging to subsystem $\mc{A}_2$ so that $\mathds{P}^{T_2} = \mathds{R}_{\mc{A}_2} \cdot \mathds{P} \cdot \mathds{R}_{\mc{A}_2}$, with $\mathds{R}_{\mc{A}_2}$ the $N_{\mc{A}}\times N_{\mc{A}}$ diagonal matrix with entries $+1$ ($-1$) corresponding to sites in $\mc{A}_1$ ($\mc{A}_2$). It can then be shown that (see, e.g., Ref.~\onlinecite{CalabreseCardy2013})
\begin{equation}
    \mc{E} = \sum_{i = 1}^{N_\mc{A}} \ln\max[1, (2\nu_i)^{-1}]
        \, .
\end{equation}
At finite temperature, a short calculation gives the elements of the correlation matrices
\begin{align}
    \mathds{Q}_{ij} &= \frac{1}{N} \sum_{\mathbf{k} \in \text{BZ}} \frac{1}{2M\omega_\mathbf{k}} \mathrm{e}^{-\mathrm{i}\mathbf{k}\vdot (\mathbf{x}_i - \mathbf{x}_j)} \coth\left(\tfrac12 \beta \omega_\mathbf{k}\right), \\
    \mathds{P}_{ij} &= \frac{1}{N} \sum_\mathbf{k \in \text{BZ}} \frac{M\omega_\mathbf{k}}{2} \mathrm{e}^{\mathrm{i}\mathbf{k}\vdot (\mathbf{x}_i - \mathbf{x}_j)} \coth\left(\tfrac12 \beta \omega_\mathbf{k}\right)
        \, ,
\end{align}
where the dispersion relation is
\begin{equation}
    \omega_{\mathbf{k}}^2 = \omega^2 + 4\varOmega^2 \sum_{\alpha=1}^2 \sin^2\left(\tfrac12 k_\alpha\right)
        \, .
\end{equation}
We have introduced the natural frequency of the nearest-neighbour interactions $\varOmega \equiv \sqrt{K/M}$, and all sums over wave vectors $\mathbf{k}$ are restricted to the first Brillouin zone (BZ).

The two Figs.~\ref{fig:harmonic_lattice_negativity} and \ref{fig:boundary_length_equal}, for the harmonic lattice and toric code respectively, share many features. Namely, $\mc{E}(T)$ remains undegraded below the gap $T \ll \omega$ in the dispersion relation, and at higher temperatures exhibits sudden death. As the length of the boundary is increased, $\mc{E}(T)$ develops more discontinuities in its first derivative and sudden death is pushed to higher temperatures. The harmonic lattice differs from the toric code in that the zero-temperature area law has subleading corrections $\sim N_\partial^{-2}$ and the number of discontinuities scales linearly, as opposed to exponentially, with $N_\partial$.

\bibliographystyle{aipnum4-1}
\bibliography{references}

\end{document}